\documentclass[aps,prl,reprint,superscriptaddress,showkeys,showpacs]{revtex4-1}

\usepackage{CJKutf8}
\usepackage{amsmath,mathtools}
\usepackage{graphicx}
\usepackage{subcaption}
\usepackage{color}
\usepackage{bm}
\usepackage[colorlinks=true,linkcolor=cyan]{hyperref}%
\usepackage{amsfonts}
\usepackage{lipsum}
\usepackage{extarrows}
\begin{document}


\title{A Time-Dependent Random State Approach for Large-scale Density Functional Calculations}

\author{
Weiqing Zhou
}%

\affiliation{Key Laboratory of Artificial Micro- and Nano-structures of Ministry of Education and School of Physics and Technology, Wuhan University, Wuhan 430072, China}%

\author{
Shengjun Yuan
}
\email{s.yuan@whu.edu.cn}
\affiliation{Key Laboratory of Artificial Micro- and Nano-structures of Ministry of Education and School of Physics and Technology, Wuhan University, Wuhan 430072, China}%

\date{\today}

\begin{abstract}
We develop a self-consistent first-principle method based on the density
functional theory. Physical quantities, such as the density of states,
Fermi energy and electron density are obtained using a time-dependent
random state method without diagonalization. The numerical
error for calculating either global or local variables always scales
as $1/\sqrt{SN_{e}}$, where $N_{e}$ is the number of electrons and
$S$ is the number of random states, leading to a sublinear computational
cost with the system size. In the limit of large systems, one random
state could be enough to achieve reasonable accuracy. The method's accuracy and scaling properties are derived analytically and
verified numerically in different condensed matter systems. Our time-dependent
random state approach provides a powerful strategy for large-scale density functional
calculations.
\end{abstract}

{\color{red}

}

\maketitle

First-principles calculation using the Density Functional Theory (DFT)
is one of the most powerful computational methods for multi-electron
systems and contributes extensively to physics, chemistry, and material
science. DFT theorems prove that there is a one-to-one mapping between
the ground-state wave function and the ground-state electron density
\cite{PhysRev.136.B864}. In the mid-1960s, Kohn and Sham showed that
the finding of the ground-state density could be determined by a set
of single-electron equations (Kohn-Sham equations) \cite{kohn1965self},
which is also known as KS-DFT. However, KS-DFT suffers from a size
limitation caused by diagonalization, in which the computational
cost exhibits a cubic scaling with the system size. Although many
efforts, such as iterative diagonalization schemes \cite{PhysRevB.53.12071},
preconditioned conjugate-gradient minimizations \cite{edelman1996conjugate,PhysRevB.40.12255,RevModPhys.64.1045},
and the Car-Parrinello method \cite{PhysRevLett.55.2471}, have improved
the scaling behaviour for a relatively small system, it is still hard
to handle systems of more than a few hundred or thousand atoms.

This size limitation has stimulated the development of linear-scaling
DFT \cite{goedecker1999linear,baer2013self,jay1999electronic,PhysRevLett.66.1438,baroni1992towards,li1993density,sanchez1997density,mohr2015accurate,goringe1997linear,hernandez1996linear,hine2009linear,vandevondele2005quickstep,ghosh2017sparc,jay1999electronic,soler2002siesta}.
The first attempt can be traced back to the 'divide and conquer' method
of Yang \cite{PhysRevLett.66.1438}. In 1992, Baroni and Giannozzi
also proposed an algorithm that determines the electron density directly
by using Green's function \cite{baroni1992towards}. In 1993, the
density-matrix minimization approach was proposed by Li, Numes, and
Vanderbilt \cite{li1993density}. Following these strategies, many
linear-scaling DFT codes have been developed \cite{mohr2015accurate,goringe1997linear,hernandez1996linear,hine2009linear,vandevondele2005quickstep,ghosh2017sparc,jay1999electronic}.
Chebyshev filter method is another successful attempt to reduce the
size of the effective dimension of Hilbert space, but there are other
non-linear factors dominated in large systems\cite{michaud2016rescu}.
A linear-scaling algorithm using atomic orbitals (LCAO) basis sets
\cite{sanchez1997density} can be applied to suitable systems with
clearly separated occupied and empty states \cite{soler2002siesta}.
Furthermore, the orbital-free DFT (OF-DFT) \cite{wesolowski2013recent,ligneres2005introduction}
is a linear-scaling approach that avoids complete diagonalization,
but the kinetic energy density functionals have not yet reached a
good accuracy for many elements \cite{zhou2005improving,wang2002orbital}.
 Recently, a linear-scaled DFT is realized by using
a stochastic technique in a trace formula \cite{baer2013self,chen2019overlapped,chen2019energy,chen2021stochastic,PhysRevLett.125.055002},
in which the statistical error of calculating a global variable, such
as the total energy (per electron), is reduced by the sample average
from different random orbitals. For local quantities, such as the
electron density, many stochastic samples are required to reach a
reasonable accuracy.

In this letter, we develop a self-consistent first-principle calculation
method based on DFT without any diagonalization of the Hamiltonian
matrix. The physical quantities, such as the density of states (DOS), Fermi
energy and the real-space distribution of electron density, are calculated
using the so-called time-dependent random state (TDRS) method. We
show that the numerical error of a global or local variable always
scales as $1/\sqrt{SN_{e}}$, where $N_{e}$ is the number of electrons
and $S$ is the number of random states. It leads to an overall sublinear
scaling of the computational costs, and one needs
fewer random states for larger systems. The method becomes extremely
powerful for massive quantum systems, and a calculation using one
random state is enough to achieve reasonable accuracy when $N_{e}\to\infty$.
Our time-dependent random-states DFT (rsDFT) originates from the real-space
TDRS method developed in the tight-binding calculations, with an extension
from globe variables (such as density of states \cite{PhysRevB.82.115448},
electronic and optical conductivities \cite{PhysRevB.84.035439},
polarization and screening functions \cite{PhysRevB.91.045420}, etc.),
to a local variable of electron density. It is a general strategy
for the local variable calculations and can be applied in the tight-binding
model or other physical models as well. 

\begin{figure}
\includegraphics[width=8.5cm]{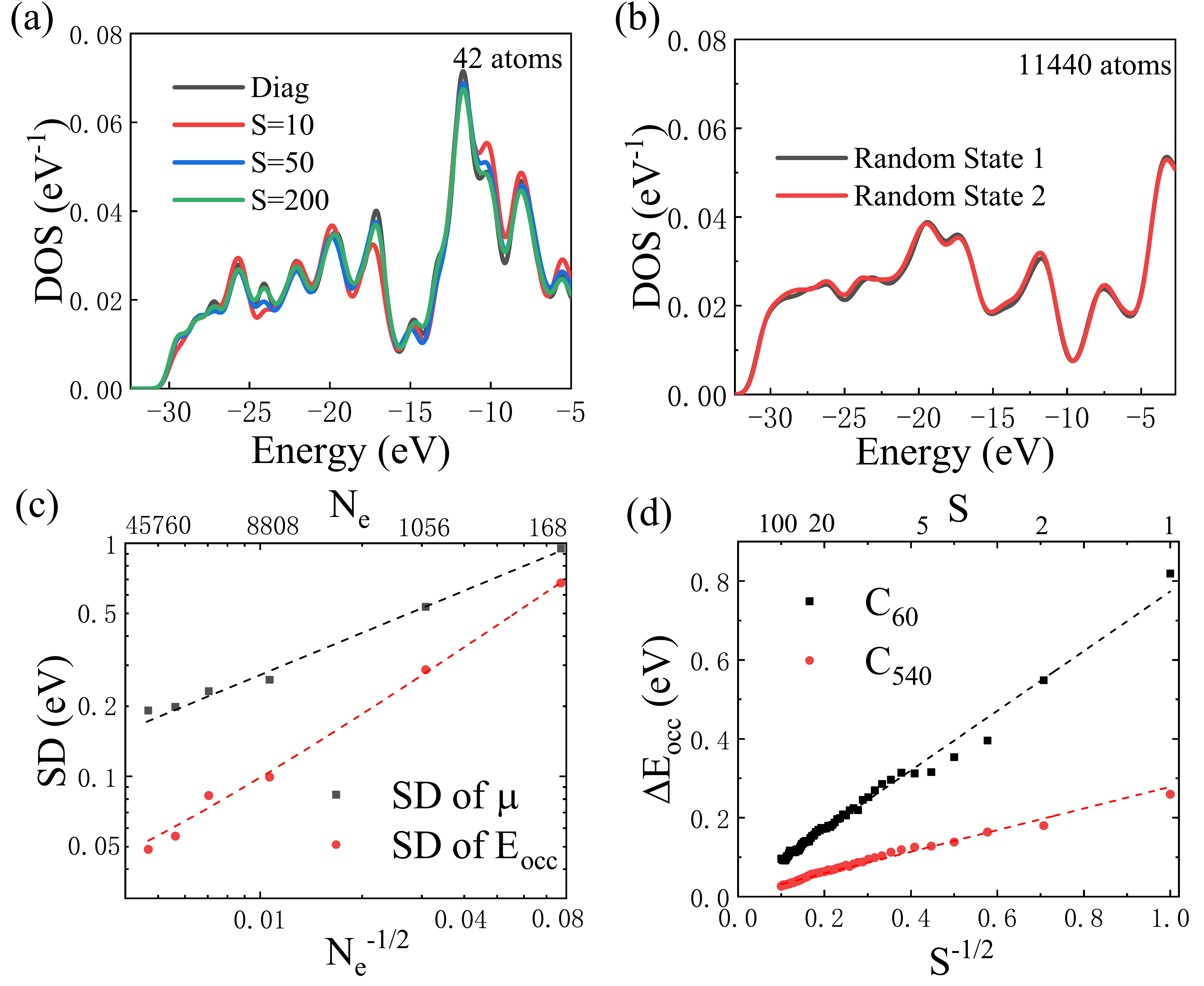} \caption{(a) The DOS of a graphite nanocrystal with 42 atoms, calculated by exact
diagonalization or using the TDRS method averaged with a different
number ($S$) of random samples. (b) The DOS of a graphite nanocrystal
with 11440 atoms was calculated using the TDRS method without random
sample averaging. The different colours represent different initial
random states. (c) The standard deviations (SD) of $\mu$ and $E_{occ}(\mu)$
for graphite nanocrystals with different size ($N_{e})$, obtained
from the statisicial analysis of results from $500$ individual random
states. (d) The error of $E_{occ}(\mu)$, with respect to the exact diagonalization,
as a function of the number of random states ($S$) used in TDRS for fullerenes
C$_{60}$ and C$_{540}$. Here, each point in (d) is averaged from
100 groups of $S$ random states.}
\label{fig:dos} 
\end{figure}

\textit{Density of states and Fermi energy.}--- In
KS-DFT, the Fermi energy is determined by counting the number of occupied
eigenstates, which are obtained from the diagonalization of the Hamiltonian.
In rsDFT, the Fermi energy is extracted by the integration of the DOS, which is calculated with the TDRS method without diagonalization \cite{hams2000fast,PhysRevB.82.115448}:
\begin{equation}
D(\varepsilon)=\frac{1}{2\pi}\int_{-\infty}^{\infty}e^{i\varepsilon t}\langle\varphi|e^{-iHt}|\varphi\rangle dt,\label{eq:dos}
\end{equation}
here, $|\varphi\rangle=\sum_{i}c_{i}|\mathbf{r_{i}}\rangle$ is a
random state in the real-space and $\{c_{i}\}$ are normalized random
complex numbers. The state $|\varphi\rangle$ is also a random superposition
state in the energy-space and can be expressed as $|\varphi\rangle=\sum_{n}b_{n}|E_{n}\rangle$,
where $b_{n}=\sum_{i}c_{i}a_{i}^{\ast}(E_{n})$ and $a_{i}(E_{n})=\langle \mathbf{r_{i}} | E_{n}\rangle$. Thus {Eq.~\ref{eq:dos},}
becomes
\begin{equation}
D(\varepsilon)=\sum_{n=1}^{N}|b_{n}|^{2}\delta(\varepsilon-E_{n}),\label{eq:dos-1}
\end{equation}
where $N$ is the dimension of the Hamiltonian. For a large but finite $N$, $|b_{n}|\rightarrow1/\sqrt{N}$, the
error of using $D(\varepsilon)$ to approximate the DOS vanish with
$1/\sqrt{N}${\cite{hams2000fast,PhysRevB.82.115448}. As we normally
use the same grid density, the dimension of the
Hamiltonian $N$, determined by the number of grids, is linearly proportional
to the number of atoms and the number of electrons. Thus the numerical error of
calculating $D(\varepsilon)$ scales also with $1/\sqrt{N_{e}}$.
The Fermi energy $\mu$ is determined by $N_{e}=\int_{-\infty}^{\mu}D(\varepsilon)d\varepsilon$.
In the case that }$N_{e}$ is not enough to provide a desired accuracy,
{additional average of $D(\varepsilon)$ from different random states
($|\varphi_{p}\rangle=\sum_{i}c_{i,p}|\mathbf{r_{i}}\rangle$, where
$p=1,2,...,S$) can be introduced to reduce the statistical error.
Then, according to the central limit theory, the overall error
of calculating $D(\varepsilon)$ scales as $1/$}$\sqrt{SN_{e}}${\cite{hams2000fast,PhysRevB.82.115448}.
Numerically, the time-evolution operator $e^{-iHt}$ can be decomposed
using the Chebyshev polynomial method as discussed in Ref.~\cite{hams2000fast,PhysRevB.82.115448},
which is unconditionally stable and leads to a linear scaling on the
system size as the Hamiltonian }$H${\ is a spare matrix in DFT.
The energy resolution is determined by $1/N_{t}\tau$, where $N_{t}$
is the number of time steps and $\tau$ is the time interval ($dt$). 

As a numerical check, we calculated the DOS of graphite nanocrystals
and fullerene with a different number of carbon atoms. Here, the Kohn-Sham
Hamiltonian is constructed with a given initial electron density $\rho(\mathbf{r})$
as
\begin{equation}
H=-\frac{\nabla^{2}}{2}+V_{ext}[\rho(\mathbf{r})]+V_{H}[\rho(\mathbf{r})]+V_{xc}[\rho(\mathbf{r})],\label{ksequation}
\end{equation}
where $-\nabla^{2}/2$ is the kinetic energy, $V_{ext}$ is the external
potential, $V_{H}$ is the Hartree potential, and $V_{xc}$ is the
exchange and correlation potential. The kinetic energy in the KS-Hamiltonian
(Eq.~(\ref{ksequation})) is approximated by using the higher-order
finite-difference expansion for the Laplacian operator in a uniform
real-space grid \cite{PhysRevLett.72.1240,chelikowsky1994higher}.
The Hartree potential $V_{H}$ is derived by solving the Poisson equation
\cite{chelikowsky1994higher}. For exchange and correlation potential
$V_{xc}$, we use the local density approximation (LDA) \cite{vosko1980accurate}.
The full ionic potential $V_{ext}$ is effectively replaced by pseudo-potential
in Kleinman-Bylander forms \cite{kleinman1982efficacious}. 

In Fig.~\ref{fig:dos}(a), we plot the DOS of a graphite nanocrystal
with 42 carbon atoms, showing that the TDRS results with more random
samples match better to the value from the diagonalization. For large
graphite nanocrystals, such as the one with 11440 atoms shown in Fig.~\ref{fig:dos}(b),
the TDRS results obtained from two individual random states are quite
close to each other. As a quantitative measurement of the statistical
error, the standard deviation (SD) of results using only one random
state are collected in Fig. \ref{fig:dos}(c). Here, in each case
we considered $500$ different random states and plotted the SDs
of $\mu$ and $E_{occ}(\mu)$ for graphite nanocrystals with different
sizes, where $E_{occ}(\mu)$ is the occupied energy defined as $E_{occ}(\mu)=\int_{-\infty}^{\mu}\varepsilon D(\varepsilon)d\varepsilon$.
It is clear that the SDs of $\mu$ and $E_{occ}(\mu)$ reduce significantly
when there are more electrons and approach to zero with an error
scales as $1/\sqrt{N_{e}}$. This indicates that for very large systems,
it is not necessary to have additional averages with different random
states, and thus using one random initial state is enough to provide
a converged result in TDRS. On the other hand, for finite systems
such as the fullerenes C$_{60}$ and C$_{540}$ shown in Fig.~\ref{fig:dos}(d), the accuracy of using TDRS can be improved by the average using more
random states and the result converges to the exact value (from diagonalization)
with an error scale as $1/\sqrt{S}$. In total, our numerical results
prove that the statistical error of using the TDRS method to calculate
DOS or related variables scales as $1/$$\sqrt{SN_{e}}$, just as
we expected (see also the error analysis of DOS in the Supplementary
Materials).

\textit{Electron density.}--- In KS-DFT, the electron
density is constructed by the superposition of occupied KS orbitals
$|E_{n}\rangle$, namely $\rho(\mathbf{r})=\sum_{n}f(E_{n})||E_{n}(\mathbf{r})\rangle|^{2}$
where $f$ is the Fermi-Dirac function. In rsDFT, the knowledge of
$|E_{n}\rangle$ is absent as we do not perform any diagonalization,
but $\rho(\mathbf{r})$ will be obtained in a different way. In the
basis of real-space grid, $\{|\mathbf{r_{i}}\rangle\}$, the wave functions
of KS orbitals can be expressed as $|E_{n}\rangle=\sum_{i=1}^{N}a_{i}(E_{n})|\mathbf{r_{i}}\rangle$,
where $N$ is the total number of grid points, i.e., the dimension
of the Hamiltonian. We consider a random state $|\varphi\rangle$
in the real space, as the one used in {Eq.~\ref{eq:dos},} which
is also a random superposition state in the energy space. Thus, a
superposition of all occupied states, with a given Fermi energy $\mu$
and temperature $T$ can be constructed by applying a Fermi-Dirac
(FD) filter on the random state as $|\varphi\rangle_{FD}\equiv\sqrt{f(H)}|\varphi\rangle=$$\sum b_{n}\sqrt{f(E_{n})}|E_{n}\rangle$,
where $f(H)=1/(e^{(H-\mu)/k_{B}T}+1)$ is the Fermi-Dirac operator.
The intensity of state vector $|\varphi\rangle_{FD}$ at grid $\mathbf{r_{j}}$
can be expressed as 
\begin{equation}
\begin{aligned} & \|\varphi\rangle_{FD}(\mathbf{r_{j}})|^{2}=\sum_{n}|b_{n}|^{2}f(E_{n})|a_{j}(E_{n})|^{2}+\\
 & \sum_{m\neq n}b_{m}^{*}b_{n}f^{\frac{1}{2}}(E_{m})f^{\frac{1}{2}}(E_{n})a_{j}^{*}(E_{m})a_{j}(E_{n}).
\end{aligned}
\label{eq:phi0phifd}
\end{equation}
As we see in the calculation of DOS, for a large but finite $N$,
$|b_{n}|\rightarrow1/\sqrt{N}$ \cite{jin2021random}, thus the first
term in {Eq.~\ref{eq:phi0phifd}} converges to $\rho(\mathbf{r_{j}})/N$,
where $\rho(\mathbf{r_{j}})=\sum_{n}f(E_{n})|a_{j}(E_{n})|^{2}$ is
exactly the electron density at grid $\boldsymbol{r}_{j}$. However,
the value of this term is of the order of $O(N_{e}/N^{2}),$which
is about $N_{e}$ times smaller than the second term $\sim O(N_{e}^{2}/N^{2})$.
This means that the second term in Eq.~\ref{eq:phi0phifd} becomes
dominant when increasing the system size. To approximate
$\rho(\mathbf{r_{j}})$ by using $\|\varphi\rangle_{FD}(\mathbf{r_{j}})|^{2}$,
one needs to reduce the second term significantly. This can be realized
by using a time-dependent approach in the following way. 

Introduce the electron density $\rho_{RS}(\mathbf{r})$ using a time-dependent RS approach as
\begin{equation}
\rho_{RS}(\mathbf{r_{j}})\equiv\frac{N}{2\pi}\int_{-\infty}^{\infty}\left\vert e^{-iHt}|\varphi\rangle_{FD}(\mathbf{r_{j}})\right\vert ^{2}dt,\label{eq:phirs}
\end{equation}
and one can prove that, $\rho_{RS}(\mathbf{r})$ converges to $\rho(\mathbf{r})$
in the limit of $N\rightarrow\infty$. Define $\delta_{\rho}\equiv\sum_{j}|\rho_{RS}(\mathbf{r_{j}})-\rho(\mathbf{r_{j}})|/N_{e}$
as a measurement error of electron density, and by using the property $\int_{-\infty}^{\infty}e^{-i(E_{n}-E_{m})t}dt=2\pi\delta(E_{n}-E_{m})$,
we have 
\begin{equation}
\begin{aligned} & \delta_{\rho}=\frac{1}{N_{e}}\sum_{j}|\sum_{n}(N|b_{n}|^{2}-1)f(E_{n})|a_{j}(E_{n})|^{2}+\\
 & 2\pi N\sum_{m\neq n}b_{m}^{*}b_{n}f^{\frac{1}{2}}(E_{m})f^{\frac{1}{2}}(E_{n})a_{j}^{*}(E_{m})a_{j}(E_{n})\delta(E_{n}-E_{m})|.
\end{aligned}
\label{eq:delta}
\end{equation}
For a large but finite $N$, $|b_{n}|\rightarrow1/\sqrt{N}$,
both terms in Eq.~\ref{eq:delta} converge to zero, but the statistical
error of the first and second terms is $O(1/\sqrt{N})$ and $O(1/N)$,
respectively. This indicates that in the limit of $N\rightarrow\infty$,
$\rho_{RS}(\mathbf{r})$ converges to $\rho(\mathbf{r})$, and $\delta_{\rho}$
is dominated by the first term and reduces to zero with a statistical
error $\sim O(1/\sqrt{N})$.

\begin{figure}
\includegraphics[width=8.5cm]{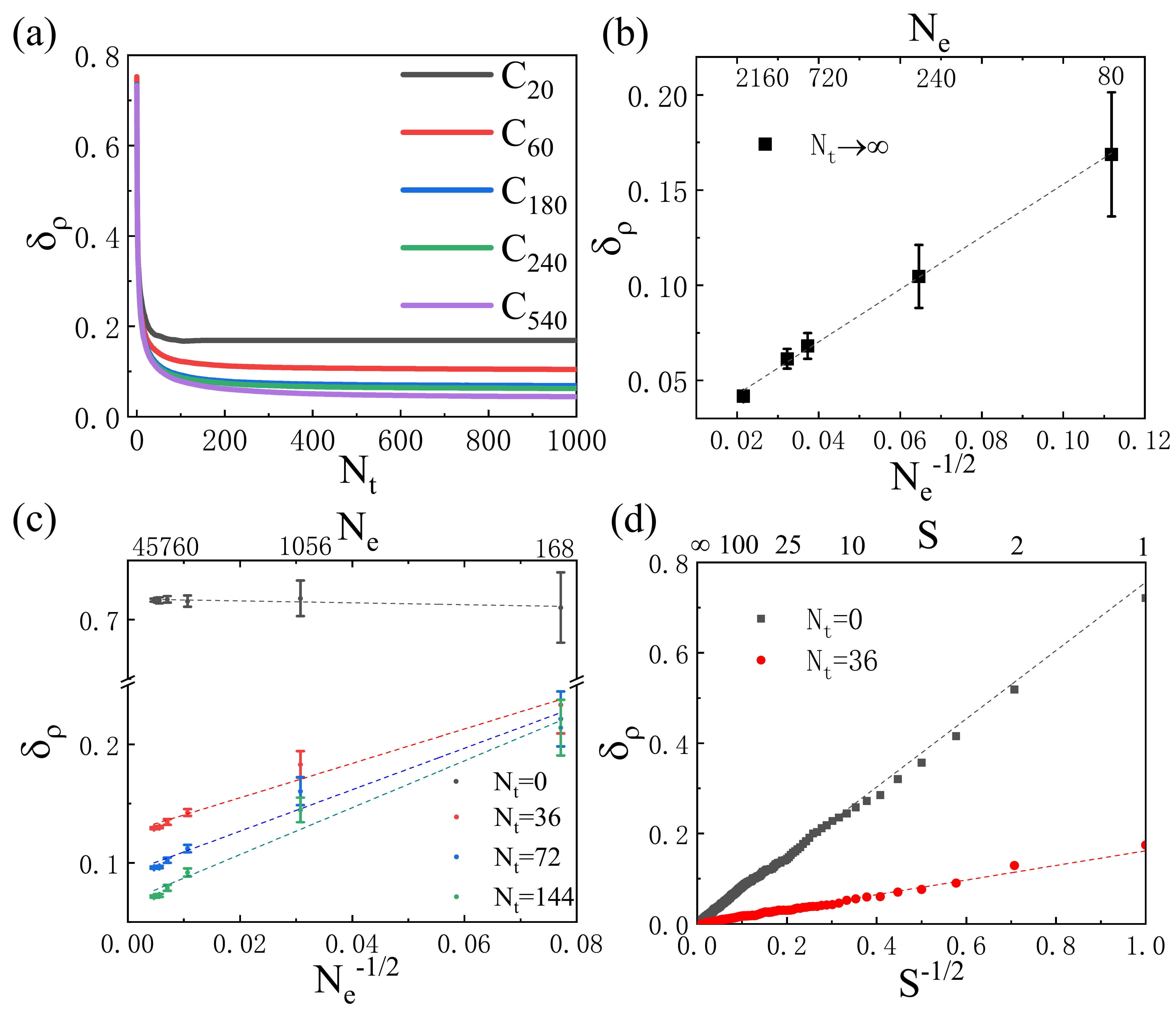} \caption{(a) The statistical error $\delta_{\rho}$ as a function of the evolution
time $N_{t}$ for C$_{20}$, C$_{60}$, C$_{180}$, C$_{240}$ and
C$_{540}$, where $\rho_{Ref}$ is the charge density obtained from
diagonalization. The time step is $\tau=64\pi$ and the number of
random samples $S=20$. (b) The values of $\delta_{\rho}$ in (a) in the limit of $N_{t}\rightarrow\infty$, plotted as a function of $N_{e}$. The error bars indicate
the corresponding standard deviations. (c) is the same as (b) but with finite
$N_{t}$, and the reference charge density is the mean value averaged from 20 random samples with $N_{t}=144$. $N_{t}=0$ means no time-dependent approach is involved
in calculating charge density. (d) The value of $\delta_{\rho}$, 
plotted as a function of sample number $S$ for C$_{60}$, without
or with time-dependent approach ($\tau=64\pi$). }
\label{fig:te} 
\end{figure}

The accuracy of using {Eq.~\ref{eq:phirs}} to obtain the electron
density can be further improved by averaging $\rho_{RS}(\mathbf{r})$
from different initial random states. Similar to the calculation of
DOS, consider a set of random states $|\varphi_{p}\rangle=\sum_{i}c_{i,p}|\mathbf{r_{i}}\rangle$,
according to the central limit theorem, for a large but finite $S$,
$\sum_{p=1}^{S}c_{i,p}c_{i',p}/S=E(|c|^{2})\delta_{i,i^{\prime}}+O(1/\sqrt{S})$\cite{hams2000fast},
where $E(|c|^{2})\sim1/N$ is the expectation value of $|c_{i}|^{2}$.
As $b_{n}=\sum_{i=1}^{N}c_{i}a_{i}^{\ast}(E_{n})$, using the normalization
property $\sum_{i=1}^{N}|a_{i}(E_{n})|^{2}=1$ and the orthogonal
property $\sum_{i=1}^{N}a_{i}(E_{n})a_{i}^{*}(E_{m})=0$ for $n\neq m$,
we have $\sum_{p}b_{m,p}^{*}b_{n,p}/S=\delta_{m,n}/N+O(1/\sqrt{S})$.
Thus, together with extra random states average, $\rho_{RS}(\mathbf{r})$
is an accurate approximation of $\rho(\mathbf{r})$ with a statistical
error $\delta_{\rho}$ scales as $1/$$\sqrt{SN_{e}}$. This scaling
behaviour is indeed the same as the calculations of DOS given in
{Eq.~\ref{eq:dos}}.

Here, the time-evolution operator $e^{-iHt}$ and the Fermi-Dirac
filter $\sqrt{f(H)}$ are carried out numerically using the Chebyshev
polynomials\cite{PhysRevB.82.115448}, which is very efficient
and accurate for sparse matrix $H$ as we discussed in the part of
DOS. In the Chebyshev decomposition of the Fermi-Dirac filter, the temperature has to be finite, and we use $T=10K$ in all the calculations.
The method introduced here becomes more efficient at high
temperatures, in which the ground state calculations also involve
many states above the Fermi energy due to a nonzero occupation probability
given by the Fermi-Dirac distribution. In the following, we show several
examples and check the accuracy of obtaining electron density using the TDRS method introduced in Eq.~\ref{eq:phirs}.

We first consider fullerene with different numbers of carbon atoms
and calculate the charge density for a given Hamiltonian using either
the TDRS method or the standard diagonalization. In Fig.~\ref{fig:te}(a),
we plot the statistical error $\delta_{\rho}$ as a function of the
number of time steps $N_{t}$ for C$_{20}$, C$_{60}$, C$_{180}$,
C$_{240}$, and C$_{540}$, where the reference charge density $\rho$
in each case is the one obtained from the diagonalization. We see
that in all cases, $\delta_{\rho}$ drops rapidly when introducing the time-evolution,
and for a given evolution period (same $N_{t}$), $\delta_{\rho}$
is always smaller in a larger sample. In Fig.~\ref{fig:te}(b), we
plot the minimum value of $\delta_{\rho}$ as a function of the number
of electrons $N_{e}$ in the limit of $N_{t}\rightarrow\infty$. It
shows clearly that $\delta_{\rho}$ scales exactly as $O(1/\sqrt{N_{e}}),$
same as we predicted from Eq.~\ref{eq:delta}. The error bars are
the standard deviations of electron density obtained from each individual
random initial state, i.e., without any random sample averaging. In
practice, it might be numerically expensive to perform a very long
evolution, and thus one needs to consider using only finite or
relatively small $N_{t}$. Thus we plot in Fig.~\ref{fig:te}(c)
similar results as (b) but with finite $N_{t}$. We see that: (1)
in the absence of the time-dependent approach ($N_{t}=0$), 
the value of $\delta_{\rho}$ keeps the same amplitude, independent of the system size;
(2) when the time-evolution is introduced, the value of $\delta_{\rho}$ starts
to decrease when increasing the size of the sample ($N_{e}$). In
Fig.~\ref{fig:te}(d), we consider the influence of sample average
on the statistical error by plotting the value of $\delta_{\rho}$
as a function of sample number $S$. Here the reference charge density
is the one obtained from C$_{60}$ with exact diagonalization, and we see that the scaling of
the error follows as $1/$$\sqrt{S}$, for both cases with or without
the time-dependent approach. The results presented
in Fig.~\ref{fig:te} verified numerically that the statistical error
of calculating the charge density using the TDRS method scales as
$1/$$\sqrt{SN_{e}}$, with the same scaling behaviour as the calculation
of DOS.

\begin{figure}
\includegraphics[width=8.5cm]{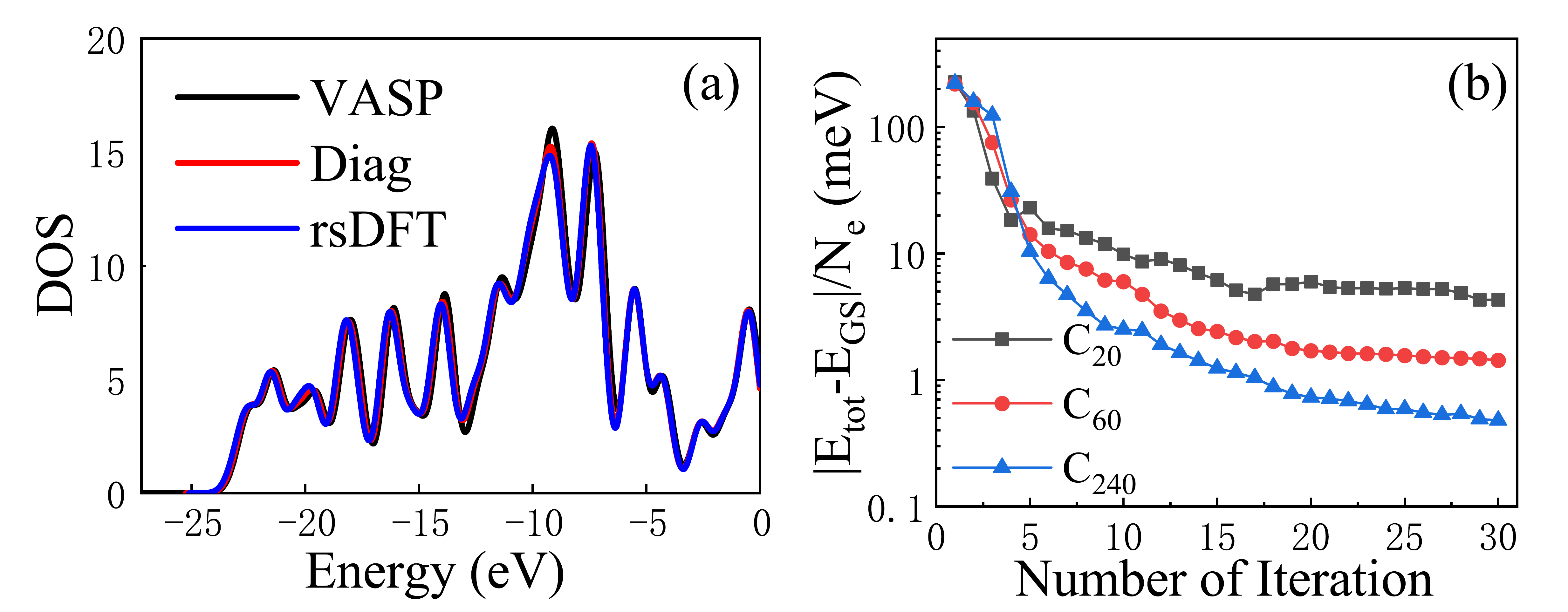} \caption{(a) For C$_{60}$, the eigenvalue distribution of the ground state
calculated from VASP, diagonalization and rsDFT, respectively. (b)The convergence of the total energy during the iteration for C$_{20}$,
C$_{60}$, and C$_{240}$. Here the total energy of the ground state ($E_{GS}$) is obtained from diagonalization.}
\label{fig:scf} 
\end{figure}

\textit{Self-consistent iteration.}--- Now, we consider the detailed self-consistent iterations in rsDFT.
Here, the numerical results obtained from the widely used commercial
KS-DFT package VASP (Vienna Ab initio Simulation Package) \cite{kresse1996efficiency}
are also presented as references.

In Fig.~\ref{fig:scf}(a), we show the ground state DOS of C$_{60}$
calculated using rsDFT, standard KS-DFT (with diagonalization) and
VASP, respectively. Pulay mix is adopted in the iteration to optimize
the input density and accelerate the convergence \cite{pulay1982improved}.
In rsDFT, we use 10 random samples in DOS calculations and 36 random
samples with $N_{t}=36,\tau=64\pi$ in electron density calculations.
The DOS obtained from different approaches agree well, indicating
that (1) our DFT code based on diagonalization correctly reproduces
the ground state from VASP, (2) the newly proposed rsDFT provides accurate
results as these can be obtained from standard KS-DFT and the diagonalization
can be completely ruled out in the entire iteration process. In Fig.~\ref{fig:scf}(b),
we present more results of converged rsDFT calculations of fullerenes
with different sizes, and show the total energy difference compared with the result from diagonalization in each
iteration step.
In general, the convergence to the ground state is more difficult
in the larger system in KS-DFT, requiring more iteration steps to
reach the same accuracy for the total energy. In rsDFT, however, the
accuracy is increased automatically in larger systems due to the $1/\sqrt{SN_{e}}$
dependence of the statistical error, as shown clearly in converge
of the total energy during the iterations in Fig.~\ref{fig:scf}(b).

\textit{Scaling Behavior.}--- At last, we check the scaling behaviour
of rsDFT. The non-diagonal elements of KS-Hamiltonian are from the
kinetic energy and non-local pseudopotentials in Eq.~(\ref{ksequation}),
leading to a highly sparse Hamiltonian matrix due to the locality.
The number of the non-zero elements in the matrix scales linearly
with the number of atoms in the system. In rsDFT, the basic and 
dominant calculations are the multiplications between the Hamiltonian
matrix and a state vector, in which the number of operations scales
linearly with the nonzero elements in the matrix. Therefore, the total
computational load (CPU time) and memory cost of rsDFT are linearly
dependent on the dimension of the Hamiltonian, which is proportional
to the number of atoms in the system. These linear-scaling behaviours
are verified using graphite crystal, with up to 11440 carbon atoms (45760 electrons)
in Fig.~\ref{scaling} (a-b). As benchmark tests, we perform complete
ground state calculations of graphite with 5040 and 11440 carbon atoms,
respectively. The difference of input and output electron density ($\Delta_{\rho}\equiv\sum_{i}|\rho_{out}(\mathbf{r_{i}})-\rho_{in}(\mathbf{r_{i}})|/N$)
as a function of iterative steps is plotted in Fig.~\ref{scaling}
(c-d). In all cases, we used the same calculation parameters, such
as the same real-space grid density, and the same accuracy for the
Chebyshev decompositions of the time-evolution operator and the Fermi-Dirac
filter. One should notice that, for 11440 carbon atoms, the total
memory cost is only 16 GB. Due to the linear-scaling behaviour, a
self-consistent calculation of millions of atoms should be possible
for computer clusters with Terabytes (TB) memory.

\begin{figure}
\includegraphics[width=8.5cm]{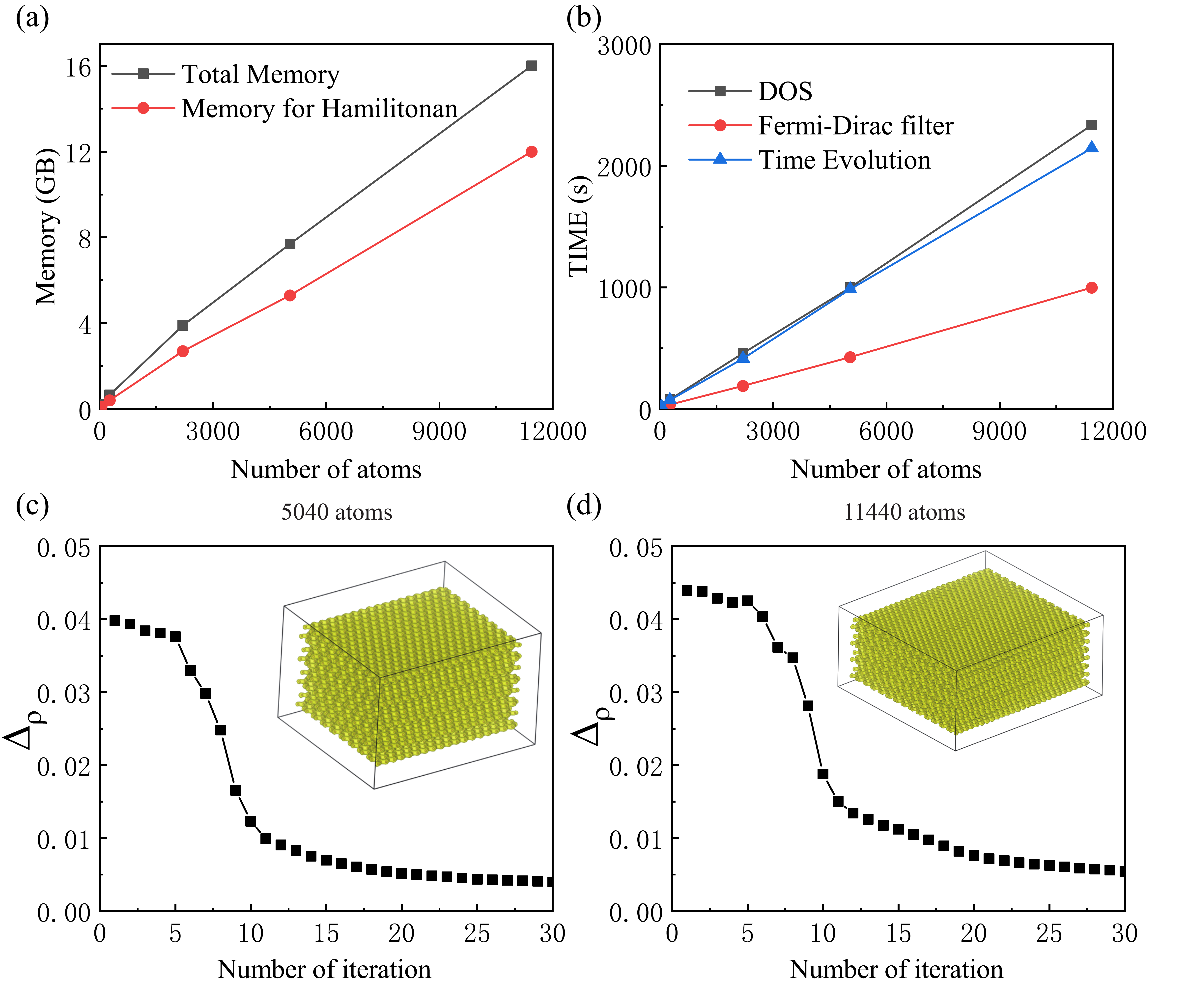} \caption{(a,b) The cost of memory and CPU time in rsDFT for A-B stacked graphite
with different numbers of carbon atoms. The black and red curves in
(a) are the total memory cost and the memory used to store the sparse
Hamiltonian matrix. The black, red and blue curves in (b) correspond
to the time cost of one iteration for the calculations of DOS, Fermi-Dirac
filter and time-evolution averaging, respectively. (c,d) The difference
of input and output electron density as a function of iteration steps
for A-B stacked graphite with 5040 and 11,440 atoms, respectively.
The insets indicate the converged ground state density.}
\label{scaling} 
\end{figure}

\textit{Conclusion and Discussion.}--- We developed a sublinear-scaled
self-consistent first-principle calculation method rsDFT. We use a
TDRS method to calculate the density of states and determine the Fermi
energy. A Fermi-Dirac filter on a random state and, subsequently,
a wave propagation according to the time-dependent Schrödinger equation
are introduced to approximate the spatial distribution of the electron
density. The accuracy can be improved by the average using different
initial random states. The overall numerical error of either a global
quantity or a local variable scales as $1/\sqrt{SN_{e}}$, where $N_{e}$
is the number of electrons and $S$ is the number of random states. It
leads to an overall sublinear scaling of the computational costs,
as for larger systems, one needs fewer random states for the sample
average. The method becomes extremely powerful when $N_{e}\to\infty$,
and a calculation using one random state is enough to achieve reasonable
accuracy. 

In the recently developed stochastic DFT (sDFT) \cite{baer2013self,chen2019overlapped,chen2019energy,chen2021stochastic,PhysRevLett.125.055002},
the physical quantities, such as the Fermi energy and electron density,
are calculated using the trace formula of the stochastic technique without
diagonalization, i.e., the trace of a variable is approximated by
the average of its expectation values in stochastic orbitals. One
of the main differences between rsDFT and sDFT is that all the variables
in rsDFT are obtained in a more deterministic way based on numerical
solutions of the time-dependent Schrödinger equation, which is absent in sDFT. In particular, the dominant noise in the electron density induced by occupied states after the Fermi-Dirac filter is dramatically reduced by the time-dependent approach in rsDFT. A large number of stochastic orbitals (random samples) is required to reduce the statistical error of the electron density in sDFT. And to keep the same accuracy of calculating the local variables such as electron density, one can not use fewer stochastic orbitals for larger systems, even in the limit of $N_{e}\to\infty$. 

In our earlier works, we have developed a so-called tight-binding
propagation method (TBPM) for large-scale modelling of complex quantum
systems. A direct extension of TBPM in rsDFT is straightforward.
For example, by using the TDRS-based method without diagonalization,
one can calculate the electronic and optical conductivities \cite{PhysRevLett.114.047403,PhysRevB.82.235409},
polarization and screening functions \cite{PhysRevLett.109.156601,PhysRevB.84.035439},
diffusion coefficient and localization length \cite{PhysRevLett.109.156601,PhysRevB.91.045420},
quasieigenstate \cite{shi2020large}, and many other applications
as implemented in our homemade simulation package, TBPLaS \cite{li2022tbplas}.
The main advantage of rsDFT and the other time-dependent methods mentioned
above is that there is no diagonalization of the Hamiltonian matrix
in the whole process, and the errors of these calculated variables
all scale as $1/\sqrt{SN_{e}}$, leading to an overall sublinear scaling on the computational costs. 
Furthermore, the atomic force can be calculated the same way as the traditional DFT or OF-DFT by using the formulation based on the electron density in real space \cite{shao2018large,ho2008introducing,golub2020conundrum,arnon2017equilibrium}. It is also possible to extract the atomic force using the wave function of approximated ground states, which will be discussed in future work.
The rsDFT provides a new possibility to
study large-scale systems from the first-principle calculations and
can be used widely in physics, chemistry, biology and material science, with possible extension to large-scale TD-DFT and GW calculations.

\textit{Acknowledgements.}--- S.Y. thanks Shiwu Gao, Hans De Raedt,
Mikhail Katsnelson, Guodong Yu, and Yalei Zhang for many helpful discussions.
This work is supported by the National Nature Science Foundation of
China (No. 11974263) and the Supercomputing Center of Wuhan University.

\bibliography{references}

\end{document}



\title{Supplementary Materials: A Time-Dependent Random State Approach for Large-scale Density Functional Calculations}

\author{Weiqing Zhou}%
\affiliation{Key Laboratory of Artificial Micro- and Nano-structures of Ministry of Education and School of Physics and Technology, Wuhan University, Wuhan 430072, China}%

\author{Shengjun Yuan}
\email{s.yuan@whu.edu.cn}
\affiliation{Key Laboratory of Artificial Micro- and Nano-structures of Ministry of Education and School of Physics and Technology, Wuhan University, Wuhan 430072, China}%


\maketitle



\subsection{1. Higher-order Finite-difference Pseudopotential Method}
Within the non-relativistic Kohn$-$Sham DFT, the ground state of a system of $N_{e}$ electrons subject to an external potential can be obtained by solving a set of one-particle equations, the Kohn$-$Sham equations (atomic units will be used throughout):
\begin{equation}
[-\frac{\nabla^{2}}{2}+V_{KS}[\rho(\mathbf{r})]]\varphi_{i}(\mathbf{r})
= \varepsilon_{i}\varphi_{i}(\mathbf{r})
\label{ksequation}
\end{equation}
where Kohn-Sham potential $V_{KS}[\rho(\mathbf{r})]$ is usually divided as:
\begin{equation}
V_{KS}[\rho(\mathbf{r})] = V_{ext}[\rho(\mathbf{r})] + V_{H}[\rho(\mathbf{r})]
+ V_{xc}[\rho(\mathbf{r})]
\end{equation}
where $V_{ext}$ is the external potential, $V_{H}$ is the Hartree potential, and $V_{xc}$ is the exchange and correlation potential. In this paper, we implement real-space finite-element methods, resulting in $V_{KS}[\rho(\mathbf{r})] = V_{KS}(\mathbf{r})$.

In our letter, we impose a simple, uniform orthogonal three-dimensional (3D) grid where the points are described in a finite domain by $(x_{i},y_{j},z_{k})$ \cite{PhysRevLett.72.1240}. Kinetic-energy operator can be described by high-order finite-element difference method \cite{chelikowsky1994higher},
\begin{equation}
\frac{\partial^{2} \varphi}{\partial x^{2}}=\sum_{n=-N_{h}}^{N_{h}} C_{n} \varphi\left(x_{i}+n h, y_{j}, z_{k}\right)+O\left(h^{2 N_{h}+2}\right)
\end{equation}
where $h$ is the grid spacing and $N_{h}$ is the order of finite-element difference. Expansion coefficients $C_{n}$ for a uniform grid are given in Table.~\ref{table:label1} \cite{chelikowsky1994higher}.
\begin{table*}[htb]
	\caption{Expansion coefficients $C_{n}$ for higher-order finite-difference expressions of the second derivative.}
	\setlength{\tabcolsep}{6mm}{
		\begin{tabular}{lcccccccc}
			\hline & $C_{i}$ &$C_{i \pm 1}$ & $C_{i \pm 2}$ & $C_{i \pm 3}$ & $C_{i \pm 4}$ & $C_{i \pm 5}$ & $C_{i \pm 6}$ \\
			\hline $N_{h}=1$ & -2 & 1 & & & & & \\
			$N_{h}=2$ & $-\frac{5}{2}$ & $\frac{4}{3}$ & $-\frac{1}{12}$ & & & & \\
			$N_{h}=3$ & $-\frac{49}{18}$ & $\frac{3}{2}$ & $-\frac{3}{20}$ & $\frac{1}{90}$ & & & \\
			$N_{h}=4$ & $-\frac{205}{72}$ & $\frac{8}{5}$ & $-\frac{1}{5}$ & $\frac{8}{315}$ & $-\frac{1}{560}$ & & \\
			$N_{h}=5$ & $-\frac{5269}{1800}$ & $\frac{5}{3}$ & $-\frac{5}{21}$ & $\frac{5}{126}$ & $-\frac{5}{1008}$ & $\frac{1}{3150}$ & \\
			$N_{h}=6$ & $-\frac{5369}{1800}$ & $\frac{12}{7}$ & $-\frac{15}{56}$ & $\frac{10}{189}$ & $-\frac{1}{112}$ & $\frac{2}{1925}$ & $-\frac{1}{16632}$ \\
			\hline \hline
		\end{tabular}
	}
	\label{table:label1}
\end{table*}

The Hartree energy density and potential are given by:
\begin{equation}
\varepsilon_{H}(\mathbf{r}) = \frac{1}{2}\int d\mathbf{r'}\frac{n(\mathbf{r'})}{|\mathbf{r}-\mathbf{r'}|}
\end{equation}
\begin{equation}
V_{H}(\mathbf{r}) = \int d\mathbf{r'}\frac{n(\mathbf{r'})}{|\mathbf{r}-\mathbf{r'}|}
\end{equation} 
The Hartree potential $V_{H}$ could be obtained by solving Poisson's equation. 

For the exchange-correlation part, we use local-density approximation (LDA):
\begin{equation}
E_{xc}[n] = \int d\mathbf{r} n(\mathbf{r})\varepsilon_{xc}(n(\mathbf{r}))
\end{equation}
The most accurate formulae for the exchange-correlation functional were obtained by fitting the QMC results for the Jellium model. Various parameterizations are available. We use one of the most popular choices proposed by Vosko-Wilk \cite{vosko1980accurate}. From the Jellium model, the local part of the exchange is given by :
\begin{equation}
\begin{aligned}
\varepsilon_{x} = -\frac{3}{4}(\frac{3}{2\pi})^{2/3}\frac{1}{r_{s}}, \\
V_{x} = (\frac{3}{2\pi})^{2/3}\frac{1}{r_{s}}.
\end{aligned}
\end{equation}

\begin{figure}
	\includegraphics[width=16cm]{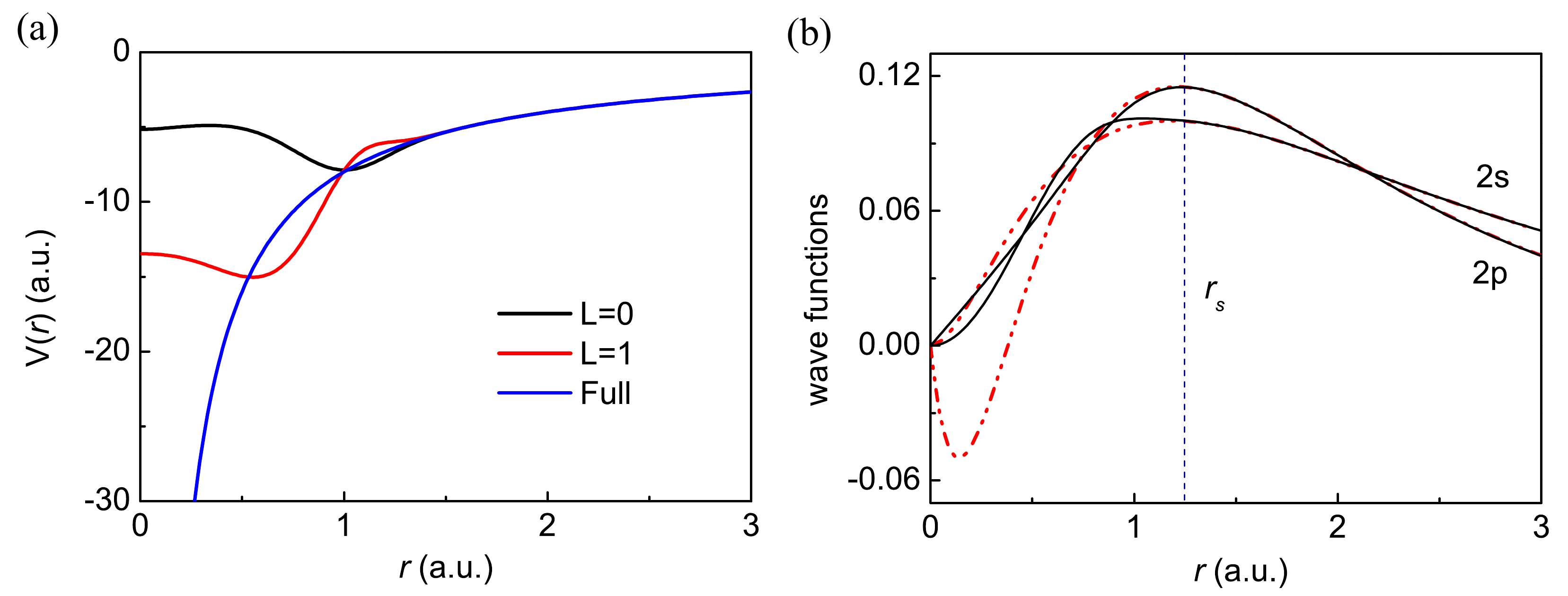}
	\caption{(a) pseudopotential of carbon atom generated by program ATOM \cite{atomcode}. (b) atomic wavefunction (Red dashed curve) and pseudo-wavefunction (black solid curve) of carbon.}
	\label{fig:cpp}
\end{figure}

For $V_{ext}(r)$, we use full ionic potential $-\frac{Z}{r}$ for the cases of single atoms. For other systems, we implement a pseudopotential operator to reduce the computational demand. We use the projection scheme of the pseudopotential operator suggested by Kleinman and Bylander \cite{kleinman1982efficacious}:
\begin{equation}
V_{\mathrm{ps}}(r)=\sum_{a=1}^{N}\left[V_{a, p s}^{l o c}(r)+\frac{\left|\Delta V_{a, p s}^{l}(r) \varphi_{l m}^{a}\right\rangle\left\langle\Delta V_{a, p s}^{l}(r) \varphi_{l m}^{a}(r)\right|}{\left\langle\varphi_{l m}^{a}(r)\left|\Delta V_{a, p s}^{l}(r)\right| \varphi_{l m}^{a}(r)\right\rangle}\right]
\end{equation}
where the total pseudopotential can be divided into non-local and local part $\Delta V_{a, \mathrm{ps}}^{l}(\boldsymbol{r}) \equiv V_{a, \mathrm{ps}}^{l}(\boldsymbol{r})-V_{a, \mathrm{ps}}^{\mathrm{loc}}(\boldsymbol{r})$. $V_{a, \mathrm{ps}}^{\mathrm{loc}}$ is the local part with specific angular momentum $l$ component of atom $a$, which differs from zero only in the region smaller than the cutoff radius $r<r_{c}$. $\varphi_{l m}^{a}$ is the atomic
pseudo wave function with $lm$ quantum angular momentum numbers. It is worth noticing that the pseudopotential operator only needs to be calculated once at the very beginning since it only depends on the atomic configuration. 
Taking the carbon atom as an example, we use the program ATOM \cite{atomcode} to generate its pseudopotential. The type of pseudopotential is chosen as local density approximation (LDA) \cite{hamann1979norm} and plotted in Fig.~\ref{fig:cpp}. In Fig.~\ref{fig:cpp} (b), we show the atomic pseudo-wave-function, and indeed it is the same as a full-potential wave-function in the range of $r>r_{s}$ where $r_{s}$ is the cutoff radius. The construction of charge density has been described in the main context.

\begin{figure}
	\includegraphics[width=16cm]{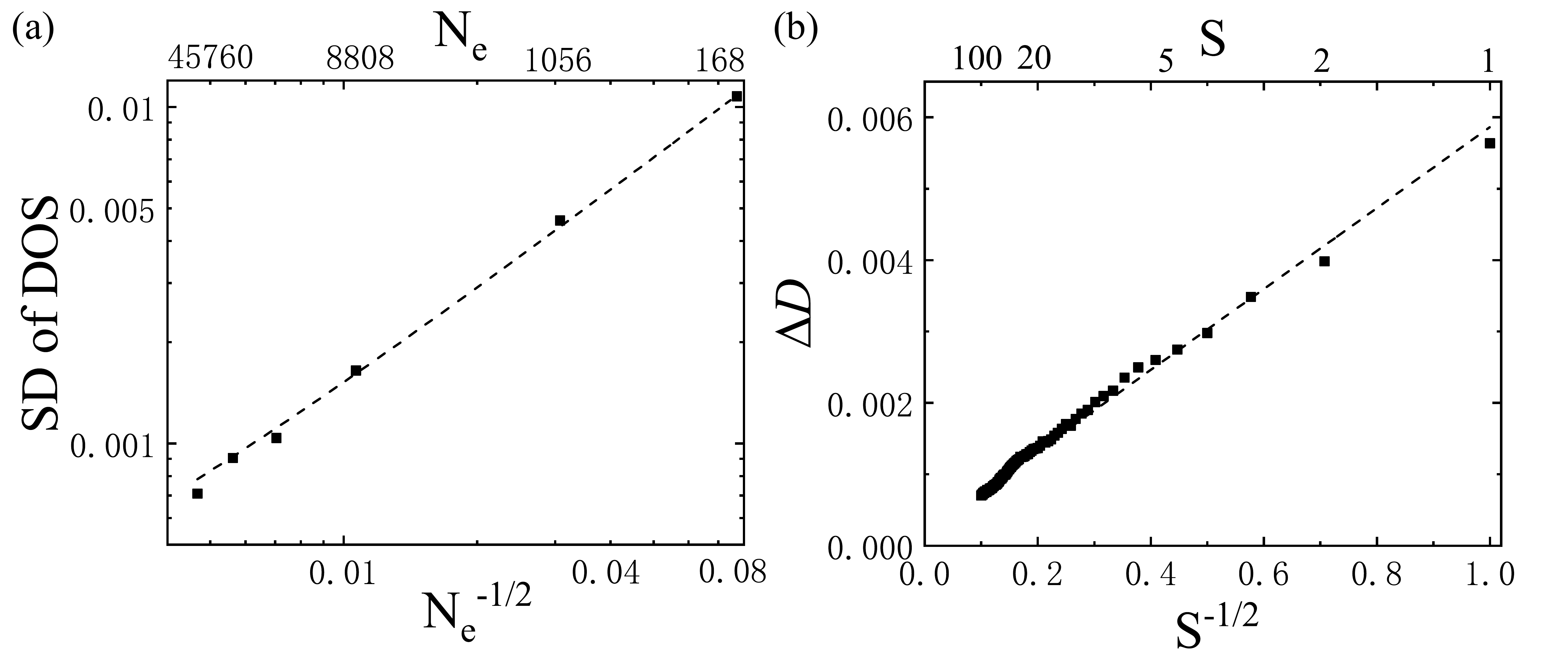}
	\caption{The statistical error of calculated DOS as a function of the number
of electrons for graphite nanocrystals (a) or the number of random
states for C$_{540}$ (b). In (a), the standard deviation of the DOS
spectrum $\int_{-\infty}^{\infty}|D(\varepsilon)-<D(\varepsilon)>|d\varepsilon$
is calculated based on the results from 500 individual random states,
where $<D(\varepsilon)>$ is the mean value of \{$D_{k}(\varepsilon)$\}
with $k=1,2,...,500.$ In (b), the error is defined as $\Delta D\equiv$$\int_{-\infty}^{\infty}|D(\varepsilon)-D_{Diag}(\varepsilon)|d\varepsilon$,
where $D_{Diag}(\varepsilon)$ is the result obtained from the diagonalization, and each point is averaged from 100 groups of $S$ random states.
}
	\label{fig:deltamu}
\end{figure}

\subsection{2. Another Fermi-Dirac Filter}
\begin{figure}
	\includegraphics[width=8cm]{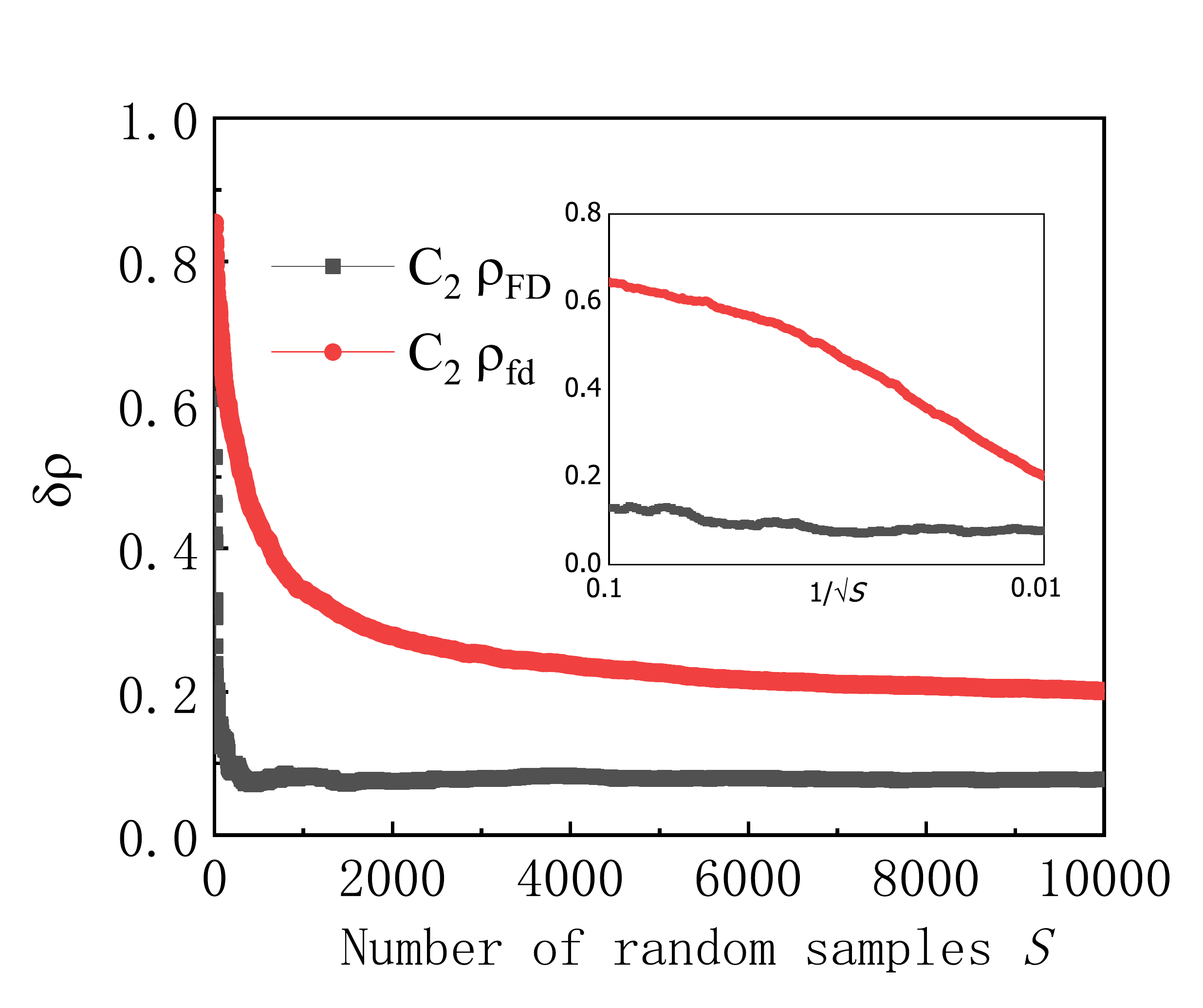}
	\caption{The difference between the electron density obtained from KS-DFT $\rho_{diag}$ and rsDFT as a function of the number of random samples $S$. In rsDFT, the results are obtained by using $\rho_{FD}$ (Eq.~(4) in the main context) or $\rho_{fd}$ (Eq.~(\ref{eq:phi0phifd})). Using $\sqrt{f(H)}$ instead of $f(H)$ in the Fermi-Dirac filter will significantly reduce the statistical error.}
	\label{fig:test_phi}
\end{figure}

In this part, we add some detailed discussion of the methods used in rsDFT. 
First, we construct a random superposition state in a uniform real-space grid as an initial state,
\begin{equation}
	|\varphi_{0}\rangle=\sum_{i=1}^{N} c_{i}|\mathbf{r_{i}}\rangle,
	\label{eq:phi0}
\end{equation}
where $N$ is the number of grid, $\{\mathbf{r_{i}}\}$ are the real space basis states, and $\{c_{i}\}$ are random complex numbers. Assuming that
\begin{equation}
    |E_{n}\rangle = \sum_{i}^{N} a_{i}(E_{n})|\mathbf{r_{i}}\rangle,
    \label{eq:ksorbital}
\end{equation}
we have 
\begin{equation}
    \begin{aligned}
    |\varphi_{0}\rangle
    &=\sum_{i=1}^{N} c_{i}\sum_{n=1}^{N}|E_{n}\rangle\langle E_{n}|\mathbf{r_{i}\rangle} \\
    & =\sum_{i=1}^{N}\sum_{n=1}^{N}c_{i}a_{i}^{*}(E_{n})|E_{n}\rangle \\
    &=\sum_{i=1}^{N}\sum_{j=1}^{N}\sum_{n=1}^{N}c_{i}a_{i}^{*}(E_{n})|\mathbf{r_{j}}\rangle \langle \mathbf{r_{j}}|E_{n}\rangle \\
    &=\sum_{i=1}^{N}\sum_{j=1}^{N}\sum_{n=1}^{N}c_{i}a_{i}^{*}(E_{n})a_{j}(E_{n})|\mathbf{r_{j}}\rangle. 
    \label{eq:phi0En}
\end{aligned}
\end{equation}
Now we consider another type of Dirac-Fermi filter different from the one introduced in the main text:
\begin{equation}
\begin{aligned}
    |\varphi\rangle_{fd}
    &\equiv f(H)|\varphi_{0}\rangle \\
    &=\sum_{i=1}^{N}\sum_{j=1}^{N}\sum_{n=1}^{N}c_{i}a_{i}^{*}(E_{n})f(E_{n})a_{j}(E_{n})|\mathbf{r_{j}}\rangle
\end{aligned}
\label{eq:phifd}
\end{equation}
In the inner product of $\langle\varphi_{0}|\varphi \rangle_{fd}$ at grid $\mathbf{r_{j}}$ can be calculated by using Eq.~(\ref{eq:phi0En}) and Eq.~(\ref{eq:phifd}),
\begin{equation}
\begin{aligned}
    &\rho_{fd}(\mathbf{r_{j}})=\prescript{}{0}{\langle} \varphi|\mathbf{r_{j}}\rangle\langle\mathbf{r_{j}}|\varphi \rangle_{fd} \\
    & = \sum_{i,i'=1}^{N}\sum_{n,m=1}^{N}c^{*}_{i}a_{i}(E_{n})a^{*}_{j}(E_{n})c_{i'}a^{*}_{i'}(E_{m})f(E_{m})a_{j}(E_{m}) \\
    & = \sum_{i,i'=1}^{N}\sum_{n=m}^{N}c^{*}_{i}a_{i}(E_{n})c_{i'}a^{*}_{i'}(E_{n})f(E_{n})|a_{j}(E_{n})|^{2} \\
    & + \sum_{i,i'=1}^{N}\sum_{n\neq m}^{N}c^{*}_{i}a_{i}(E_{n})a^{*}_{j}(E_{n})c_{i'}a^{*}_{i'}(E_{m})f(E_{m})a_{j}(E_{m})
    \label{eq:phi0phifd}
\end{aligned}
\end{equation}
According to the central limit theorem, for a large but finite number ($S$) of the random states $|\varphi_{p}\rangle=\sum_{i}c_{i,p}|\mathbf{r_{i}}\rangle$, we have
\begin{equation}
    \frac{1}{S}\sum_{p=1}^{S} c_{i,p}c_{i',p}=E(c^{2})\delta_{i,i'} + O(\frac{1}{\sqrt{S}}). 
    \label{eq:cii}
\end{equation}
Therefore, one proves that
\begin{equation}
\begin{aligned}
& \lim _{S \rightarrow \infty} \frac{1}{S}\sum_{p=1}^{S} \langle\varphi_{p}|\mathbf{r_{j}}\rangle\langle\mathbf{r_{j}}|\varphi_{p} \rangle_{fd} \\
& = \sum_{i=1}^{N}\sum_{n=1}^{N}E(|c|^{2})f(E_{n})|a_{i}(E_{n})|^{2}|a_{j}(E_{n})|^{2} \\
& + \sum_{i=1}^{N}\sum_{n\neq m}^{N}E(|c|^{2})f(E_{m})a_{i}(E_{n})a^{*}_{i}(E_{m})a^{*}_{j}(E_{n})a_{j}(E_{m}) \\
& = \sum_{i=1}^{N}|a_{i}(E_{n})|^{2}\sum_{n=1}^{N}E(|c|^{2})f(E_{n})|a_{j}(E_{n})|^{2} \\
& = \frac{1}{N}\sum_{n=1}^{N}f(E_{n})|a_{j}(E_{n})|^{2} \\
& = \frac{1}{N}\rho_{diag}(\mathbf{x_{j}})
\end{aligned}
\label{eq:rhofd1}
\end{equation}
here we used the normalization property of KS orbitals 
\begin{equation}
\sum_{i=1}^{N}|a_{i}(E_{n})|^{2}=1
\end{equation}
and the orthogonal property 
\begin{equation}
\sum_{i=1}^{N}a_{i}(E_{n})a^{*}_{i}(E_{m})=0
\end{equation}
for $m \neq n$. Eq.~(\ref{eq:rhofd1}) indicates that 
\begin{equation}
\rho_{fd}(\mathbf{r_{j}}) \equiv \frac{N}{S}\sum_{p=1}^{S} \langle\varphi_{p}|\mathbf{r_{j}}\rangle\langle\mathbf{r_{j}}|\varphi_{p} \rangle_{fd}
\end{equation}
is an approximation of the charge density at $\mathbf{r_{j}}$ with an error vanishes as $1/ \sqrt{S}$, which can be verified in the zoom-in figure of Fig.~\ref{fig:test_phi} .

However, 
the converge of $\rho_{fd}(\mathbf{r_{j}})$ using the Fermi-Dirac filter $f(H)$ is slower than the one using double $\sqrt{f(H)}$ introduced in the main text. The reason is that, in Eq.~(\ref{eq:phi0phifd}), the sum in the second term ($n \neq m$) involves all unoccupied states associated with the index $n$, and their number is several orders larger than the number of occupied states because $N \gg N_{e}$. 
To overcome this difficulty, we introduce $|\varphi\rangle_{FD}=\sqrt{f(H)}|\varphi\rangle_{0}$, the one used in the main context. The main advantage of using $\rho_{FD}(\mathbf{r_{j}})$ is that the sums in the second term ($n \neq m$) of Eq.~(4) of the main context includes only occupied states, leading to a much faster convergence compared with Eq.~(\ref{eq:phi0phifd}) (see Fig.~\ref{fig:test_phi}). 

\subsection{3. Chebyshev Polynomials Method}
In the numerical calculation, the operators $\frac{1}{\sqrt{e^{\beta(H-\mu)}+1}}$ and $e^{-iHt}$ are approximated by using the Chebyshev polynomial method. In general, a function $f(x)$ whose values are in the range [-1,1] can be expressed as,
\begin{equation}
	f(x)=\frac{1}{2} c_{0} T_{0}(x)+\sum_{k=1}^{\infty} c_{k} T_{k}(x)
\end{equation}
where $T_{k}(x)=\cos (k \arccos x)$ and the coefficients $c_{k}$ are 
\begin{equation}
	c_{k}=\frac{2}{\pi} \int_{-1}^{1} \frac{d x}{\sqrt{1-x^{2}}} f(x) T_{k}(x)
\end{equation}
if we let $x=cos\theta$, then $T_{k}(x)=T_{k}(\cos \theta)=\cos k \theta$, and
\begin{equation}
\begin{aligned}
c_{k} &=\frac{2}{\pi} \int_{0}^{\pi} f(\cos \theta) \cos k \theta d \theta \\
&=\operatorname{Re}\left[\frac{2}{N} \sum_{n=0}^{N-1} f\left(\cos \frac{2 \pi n}{N}\right) e^{2 \pi i n k / N}\right],
\end{aligned}
\label{ck}
\end{equation}
which can be calculated by the fast Fourier transform (FFT). 
We normalize $H$ such that $\widetilde{H}=H /\|H\|$ has eigenvalues in the range [-1,1] and put $\widetilde{\beta}=\beta /\|\beta\|$. Then 
\begin{equation}
	f(\tilde{H})=\sum_{k=0}^{\infty} c_{k} T_{k}(\tilde{H})
	\label{FOE}
\end{equation}
where the Chebyshev polynomial $T_{k}(x)$ is the Chebyshev polynomial of the first kind. $T_{k}(x)$ obeys the following recurrence relation:
$$T_{k+1}(x)+T_{k-1}(x)=2 x T_{k}(x)$$
with
$$T_{0}(x)=1, T_{1}(x)=x.$$

In Table.~\ref{table:label2}, we present the number of nonzero Bessel function ($N_{Bessel}$) as a function of time step $\tau$, and the corresponding number of total matrix-vector operations $N_{operations}$ for the same propagation time $T=1024\pi$. We can see a larger $\tau$ leads to fewer operations with the same total propagation time.

\begin{figure}[tbp]	
	\includegraphics[width=14cm]{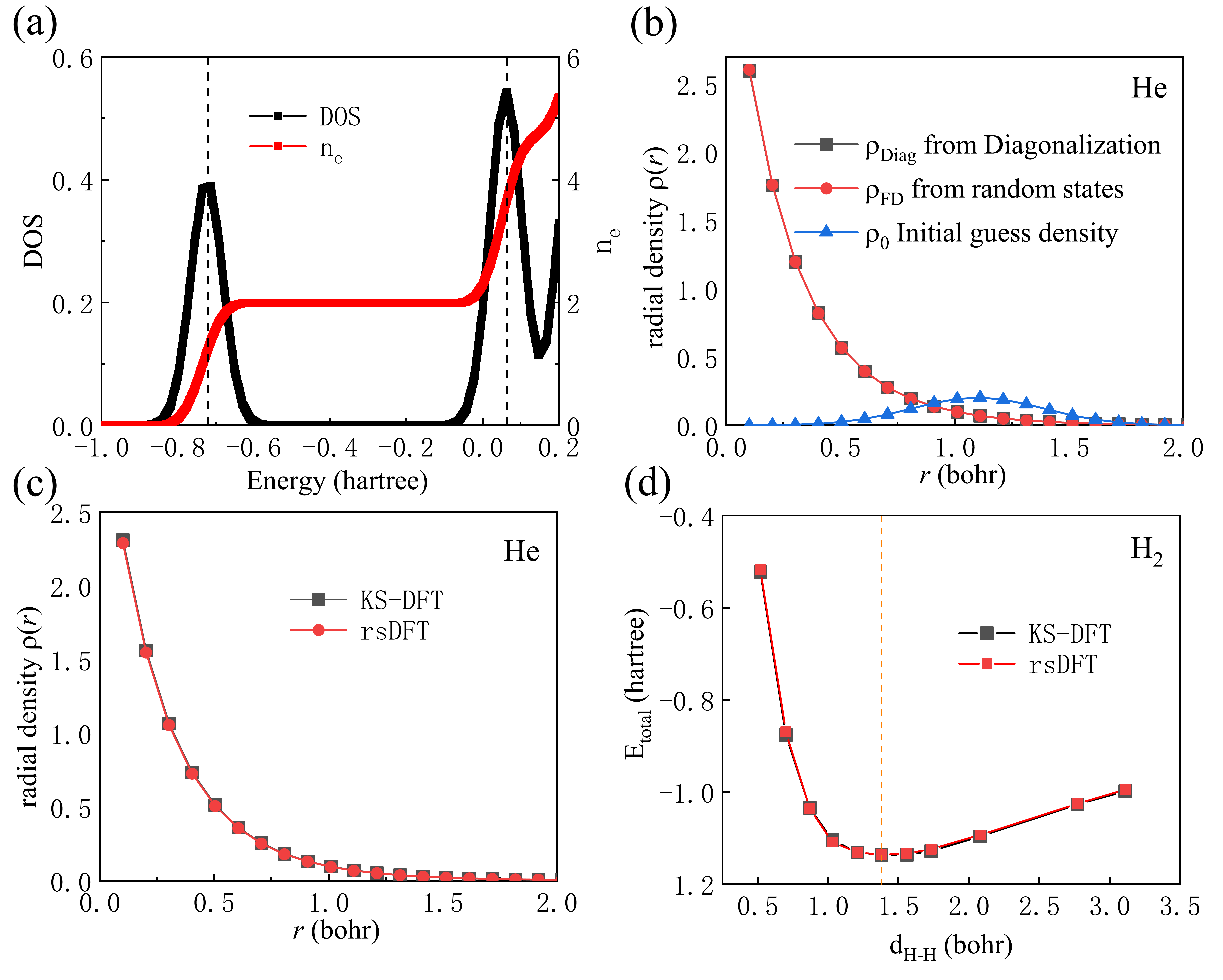}
	\caption{(a) DOS and carrier density of one He atom calculated using the time-dependent random state method of one He atom. The vertical dashed lines indicate the energies obtained from the diagonalization. (b) The output electron density after one iteration using diagonalization (black) and Fermi-Dirac filter on random states (red) from the same input electron density (blue). For the He atom, due to the spherical symmetry, the electron density is only a function of distance ($r$) from the center of the atom. (c) The converged ground-state electron density of the He atom was obtained from KS-DFT and rsDFT.  (d) The total energy as a function of bond length of molecular H$_{2}$ from obtained from KS-DFT and rsDFT, respectively.}
	\label{fig:he}
\end{figure}

\begin{figure}[tbp]	
	\includegraphics[width=12cm]{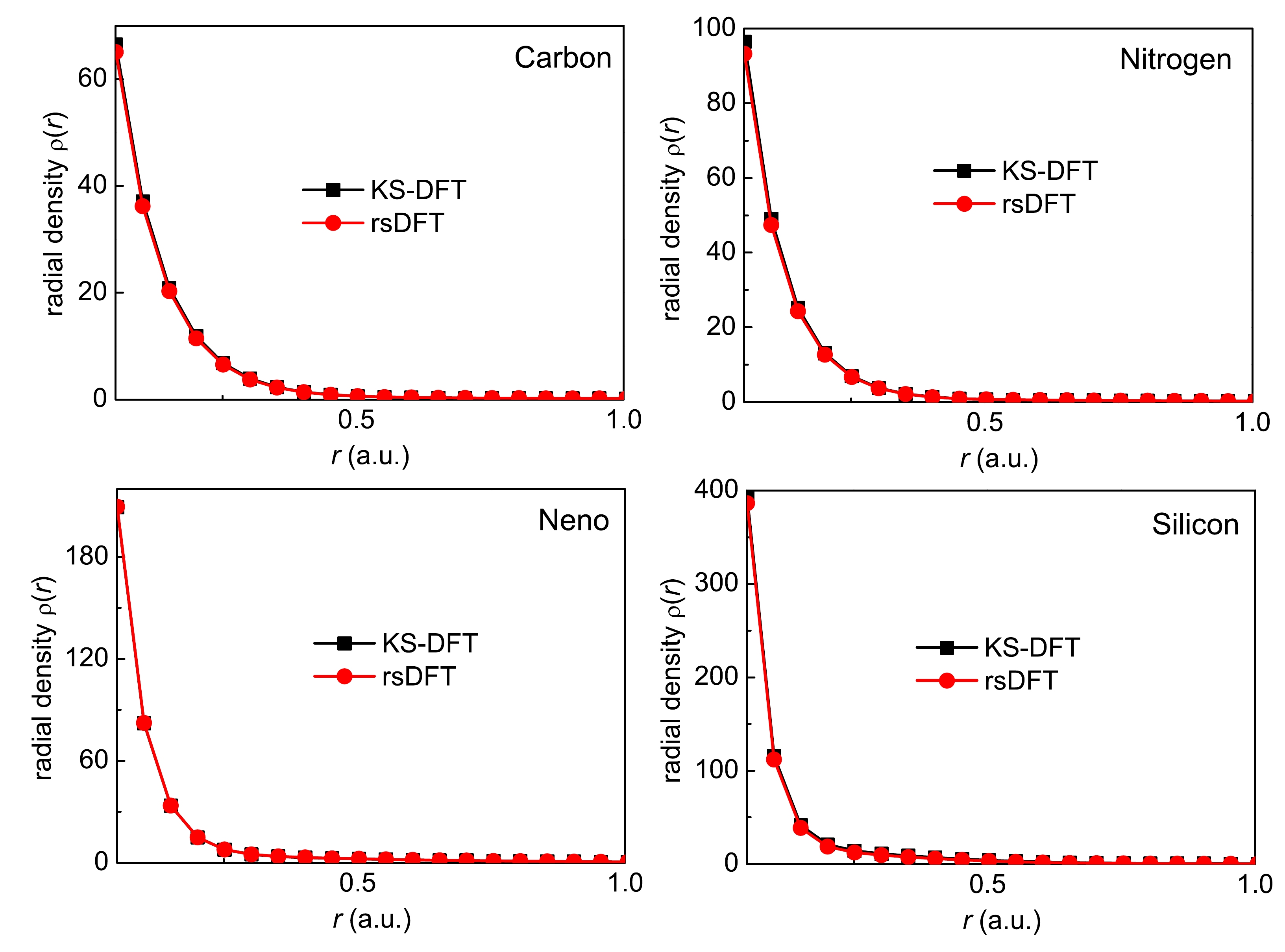}
	\caption{Comparisons of ground-state charge density calculated by KS-DFT based on diagonalization and rsDFT of single atoms.}
	\label{fig:atom}
\end{figure}
\begin{figure}
	\includegraphics[width=12cm]{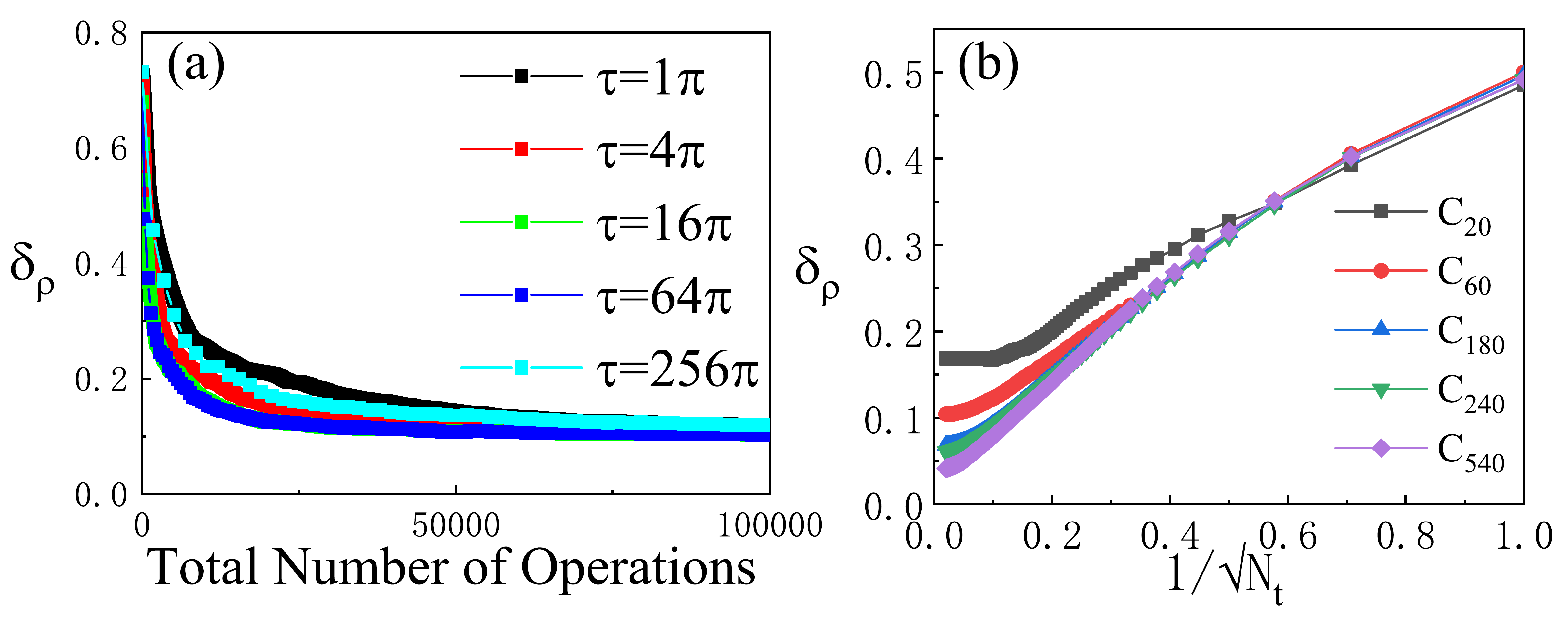}
	\caption{(a) The statistical error $\delta_{\rho}$ as a function of total operations. Different colour indicates different time step $\tau$ in time evolution. (b) The statistical error $\delta_{\rho}$ as a function of $1/\sqrt{N_{t}}$ for different fullerenes, where $\tau=64\pi$. 
 }
	\label{fig:fdte}
\end{figure}

\begingroup
\setlength{\tabcolsep}{50pt} 
\begin{table}
\caption{The number of nonzero Bessel function ($N_{Bessel}$) as a function of time step $\tau$, and the corresponding number of total matrix-vector operations $N_{operations}$ for the same certain propagation time $T=1024\pi$.}
\begin{tabular}{cccc}
\hline
$\tau$ & $N_{Bessel}$ &	$N_{t}$ & $N_{operations}$ \\
\hline
$\pi$ & 20 & 1024 & 20480 \\
$2\pi$ & 27 & 512 & 13824 \\
$4\pi$ & 38 & 256 & 9728 \\
$8\pi$ & 56 & 128 & 7168 \\
$16\pi$ & 89 & 64 & 5695 \\
$32\pi$ & 149 & 32 & 4768 \\
$64\pi$ & 261  & 16 & 4176 \\
$128\pi$ & 478 & 8 & 3824 \\
$256\pi$ & 899 & 4 & 3596 \\
$512\pi$ & 1727 & 2 & 3454 \\
\hline
\end{tabular}
\label{table:label2}
\end{table}

\begin{figure}
	\includegraphics[width=16cm]{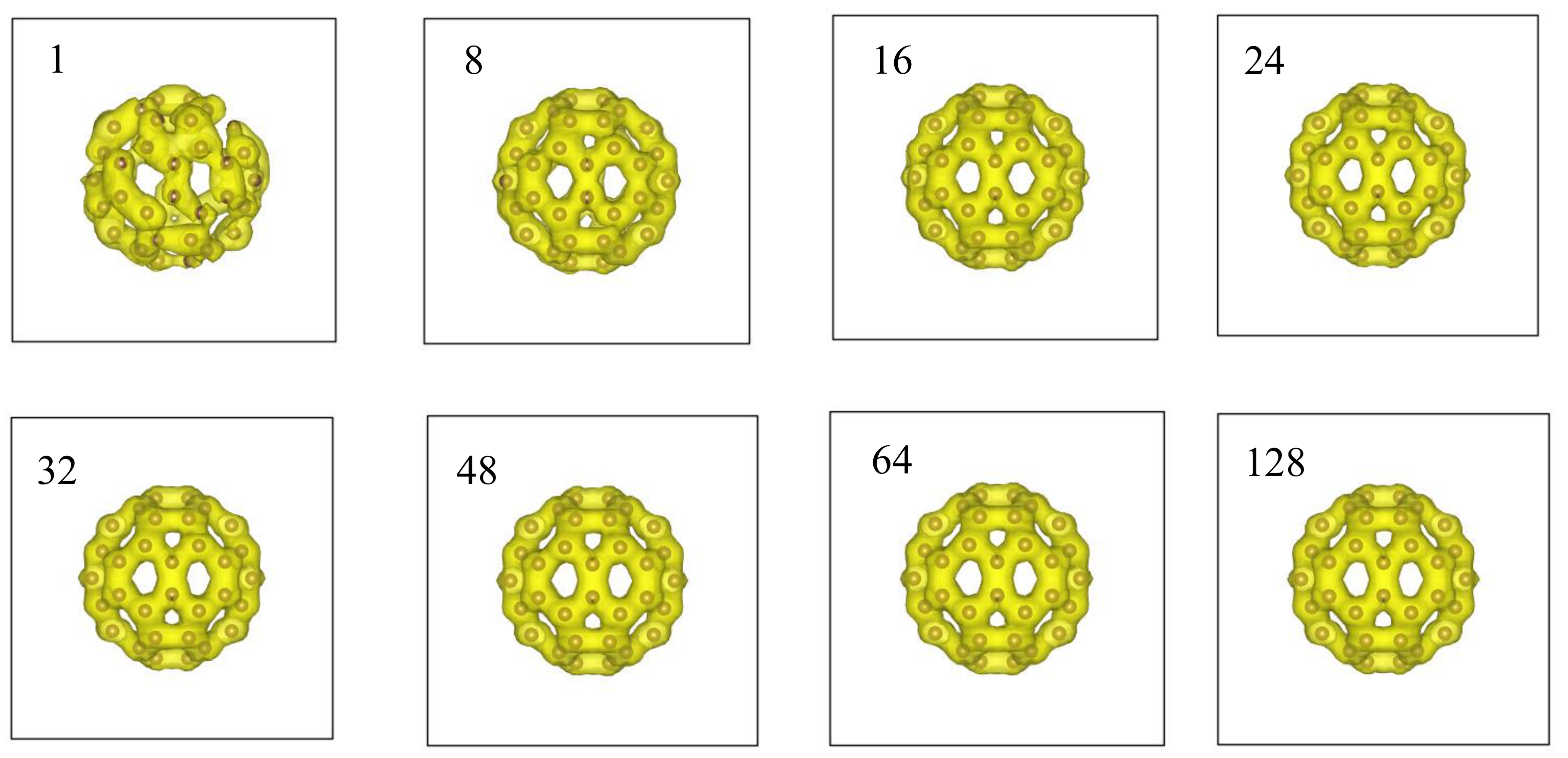}
	\caption{The charge density of C$_{60}$ calculated from average over a different number of random samples without time evolution.}
	\label{fig:C60_fd}
\end{figure}
\begin{figure}
	\includegraphics[width=16cm]{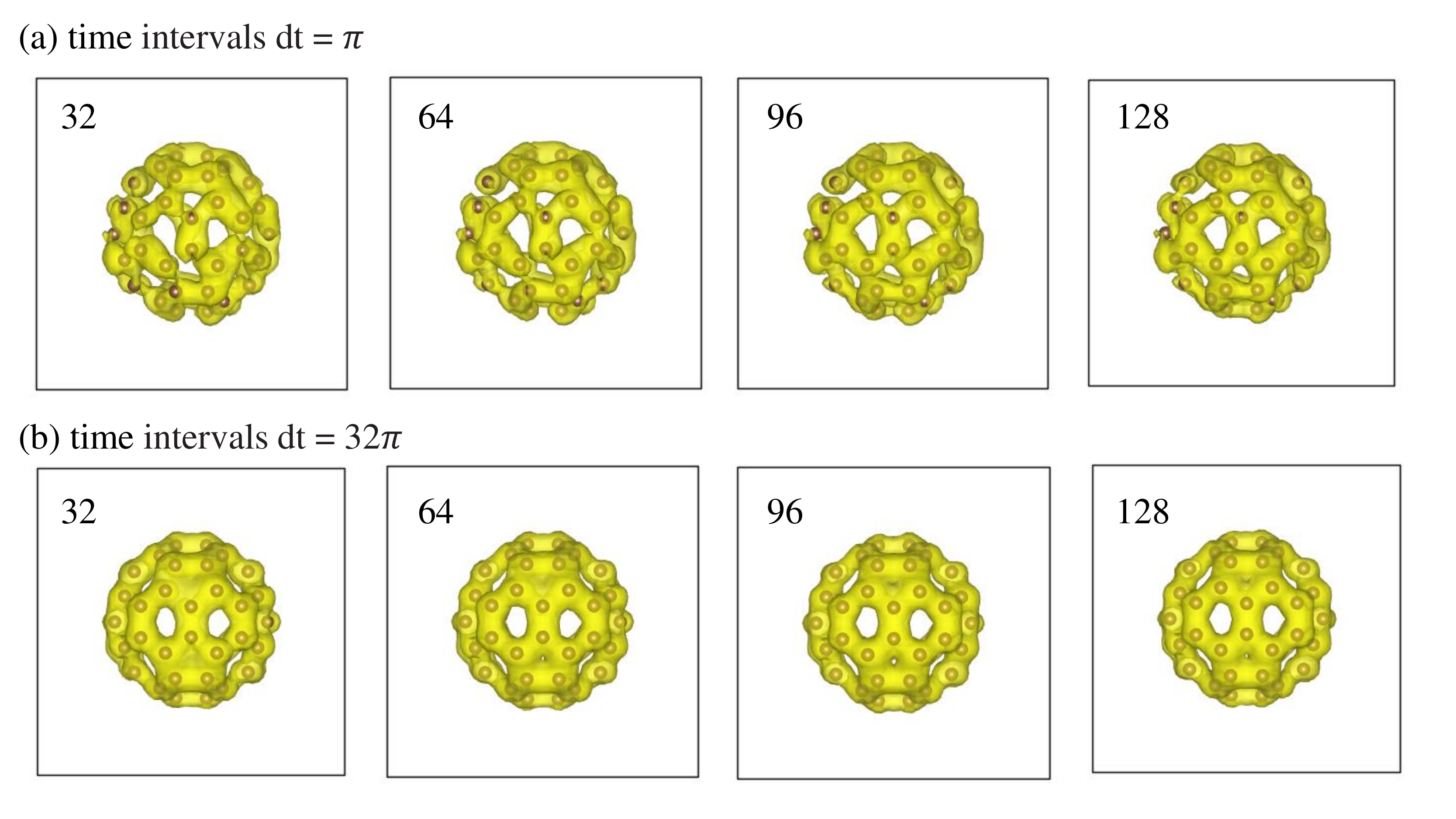}
	\caption{The charge density of C$_{60}$ calculated from one random state averaging over from time evolution for 32, 64, 96, 128 steps with dt=$\pi$ (a) and dt=$32\pi$ (b).}
	\label{fig:C60_time}
\end{figure}

\section{4. Single Atom}

Let us consider a single Helium atom.
In step (I), we construct a KS-Hamiltonian based on an initial electron density. 
In step (II), we obtain DOS $D(\varepsilon)$ by using the time-evolution method without the diagonalization of the Hamiltonian matrix and subsequently determine the Fermi level $\mu$ (see Fig.~\ref{fig:he}(a)).

As a comparison, the energies of KS orbitals from the diagonalization are also shown in Fig.~\ref{fig:he}(a), which agree very well with our results.
In step (III), as there is only one occupied state, one can just use $\rho_{FD}$ to obtain the electron density without time evolution. As a comparison, we also calculate the electron density $\rho_{diag}$ based on the occupied KS orbital obtained from the diagonalization of KS-Hamiltonian,
and plot together with $\rho_{FD}$ in Fig.~\ref{fig:he}(b). We see that with only $S$=10 random samples, $\rho_{FD}$ converges to $\rho_{diag}$ with an error of $\Delta(\rho_{FD}-\rho_{diag})=8.06\times10^{-5}$, where $\Delta(\rho_{FD}-\rho_{diag}) \equiv \sum_{j=1}^{N}{|\rho_{FD}(\mathbf{r_{j}})-\rho_{diag}(\mathbf{r_{j}})|/N}$. In step (IV), we use $\rho_{FD}$ as the new input electron density and perform the next iteration. The self-consistent iterations, including steps (I) to (IV), are continued until a threshold is reached. In our approach, since the space resolution (determined by $N$) is much larger than the energy resolution (determined by $N_{t}$), it is more accurate to use the electron density instead of the total energy to define the convergence criterion. The ground-state electron density obtained from rsDFT without any diagonalization agrees well with the one from the common KS-DFT with diagonalization. Both are plotted in Fig.~\ref{fig:he}(c) for comparison. More examples of other single atoms can be found in Fig.~\ref{fig:atom}.

\section{4. Molecules}

For molecular systems, we consider a diatomic model H$_{2}$. The iterative calculations are similar to those of a single atom. Here we verify our approach by calculating the total energies for different H-H bond lengths and compare the results from our rsDFT approach and the common KS-DFT in Fig.~\ref{fig:he}(d). The two methods yield similar total energies for a given H-H bond length. The bond lengths in the ground state obtained from both ways are the same (74 pm), which agrees with the well-known result \cite{huber2013molecular}.

\section{5. Clusters}
We extend our calculations to large atomic clusters of fullerenes C$_{60}$ and C$_{540}$.
In Fig.~\ref{fig:C60_fd}, we plot the $\rho_{FD}$ averaging from up to 128 random states. As a comparison, we also present $\rho_{RS}$ using only one random state, but different propagation time in Fig.~\ref{fig:C60_time}(a) with $\tau=\pi$ and Fig.~\ref{fig:C60_time}(b) with $\tau=32\pi$.
The real-space distribution of electron density in the ground state is visualized by VESTA \cite{momma2008vesta} in Fig.~\ref{fig:c60ground}. 
We use VASP (Vienna Ab initio Simulation Package) \cite{kresse1996efficiency} to represent the standard KS-DFT method. VASP is a very efficient and widely used commercial KS-DFT package. 
The electron density distributions obtained from rsDFT and VASP are very similar.

\begin{figure}
	\includegraphics[width=12cm]{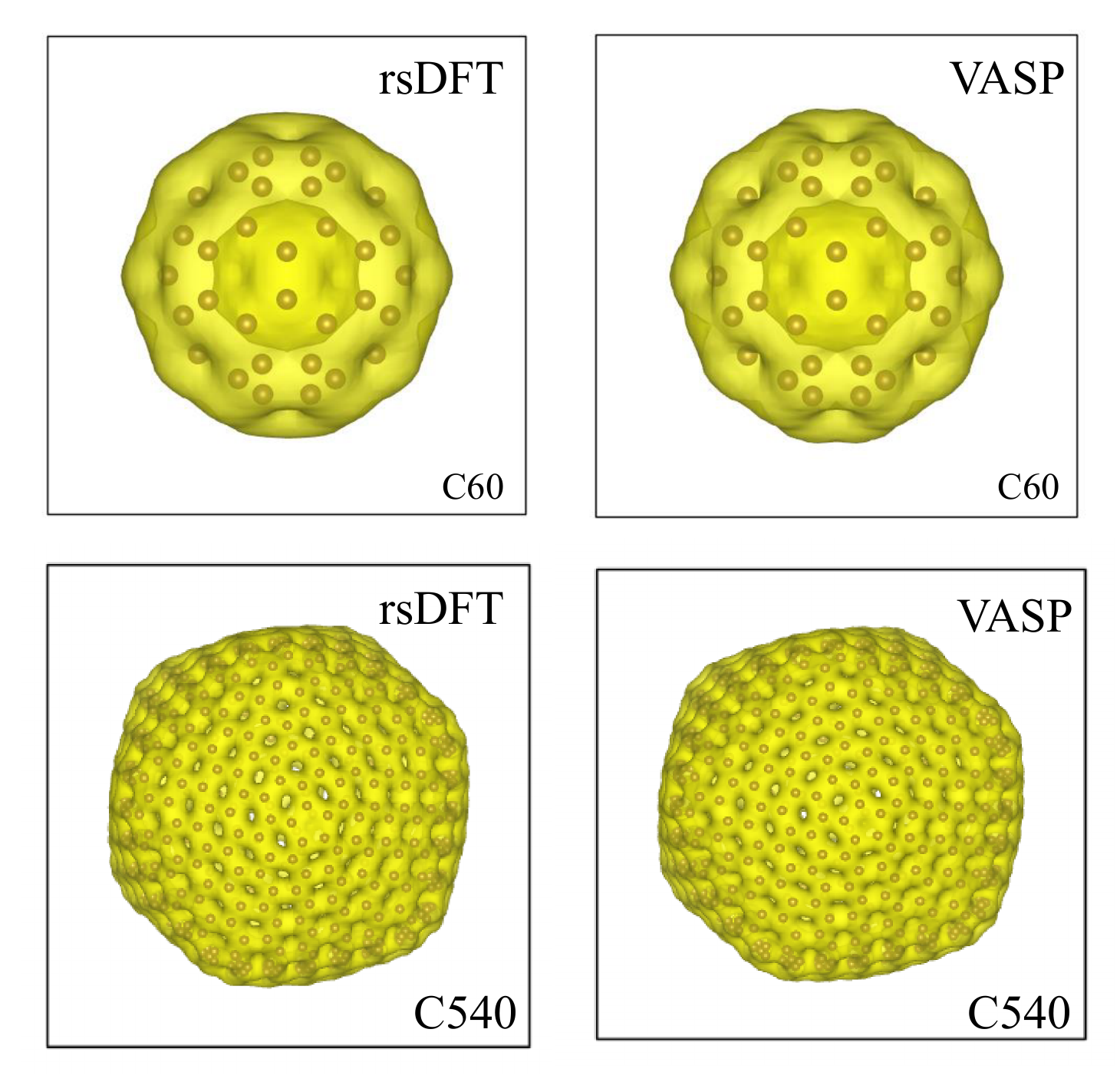}
	\caption{The ground-state charge density calculated by rsDFT and VASP, respectively.}
	\label{fig:c60ground}
\end{figure}

\section{6. Crystals}
More rsDFT calculations of graphite nanocrystals with different numbers
of carbon atoms are plotted in Fig.~\ref{fig:graphite}.

\section{7. CPU Time and Memory Cost}
To have a direct comparison of the CPU time and memory
cost between the traditional KS-DFT (with diagonalization) and rsDFT
(without diagonalization), we performed calculations for the fullerenes
with different numbers of atoms (electrons) on a server with 40 CPU
cores (2{*}Intel(R) Xeon(R) CPU Gold 6248). As shown in Fig.~\ref{fig:time}(a),
if the system has less than $\sim1000$ electrons, the traditional
KS-DFT is much faster, but when the system size reaches $\sim1000$
electrons, the rsDFT method becomes more efficient. The results in
Fig.~\ref{fig:time}(a) also indicates that the time cost of rsDFT
scales linearly with the system size, whereas the traditional KS-DFT
scales approximately as $O(N_{e}^{3})$. Although the accuracy of
rsDFT and KS-DFT are not exactly the same, we estimate that rsDFT
becomes more efficient when the system contains a few thousand or
more electrons.

\begin{figure*}
	\includegraphics[width=18cm]{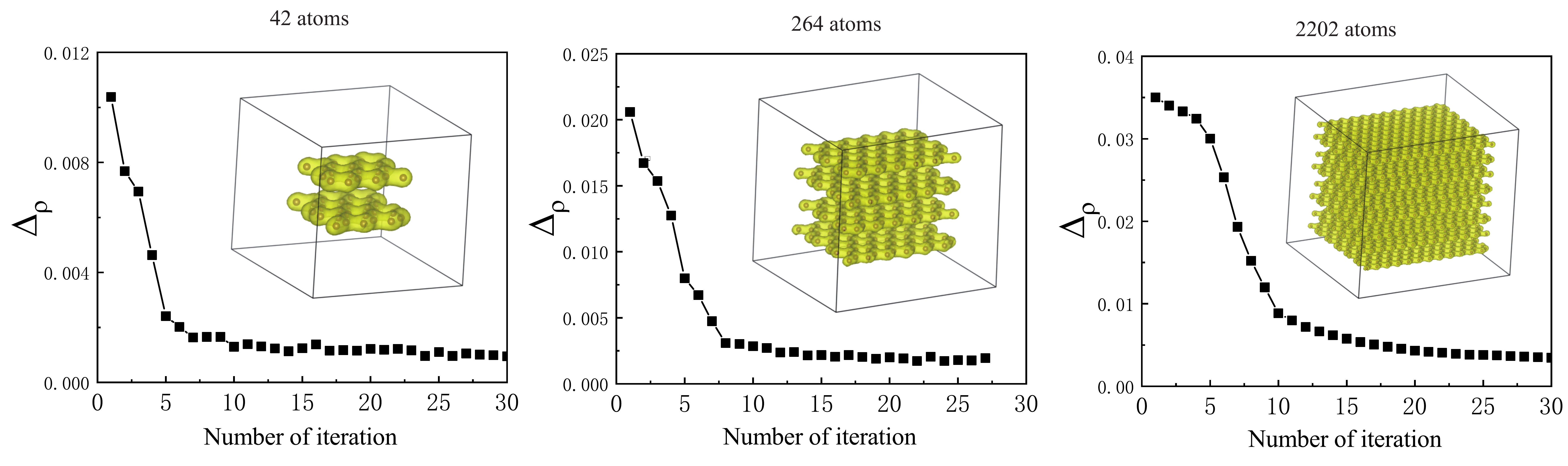}
	\caption{$\Delta(\rho_{in}-\rho_{out})$ as a function of iterative steps for A-B stacked graphite calculated by rsDFT. The insets indicate
the converged ground state densities.}
	\label{fig:graphite}
\end{figure*}

\begin{figure}
\includegraphics[width=14cm]{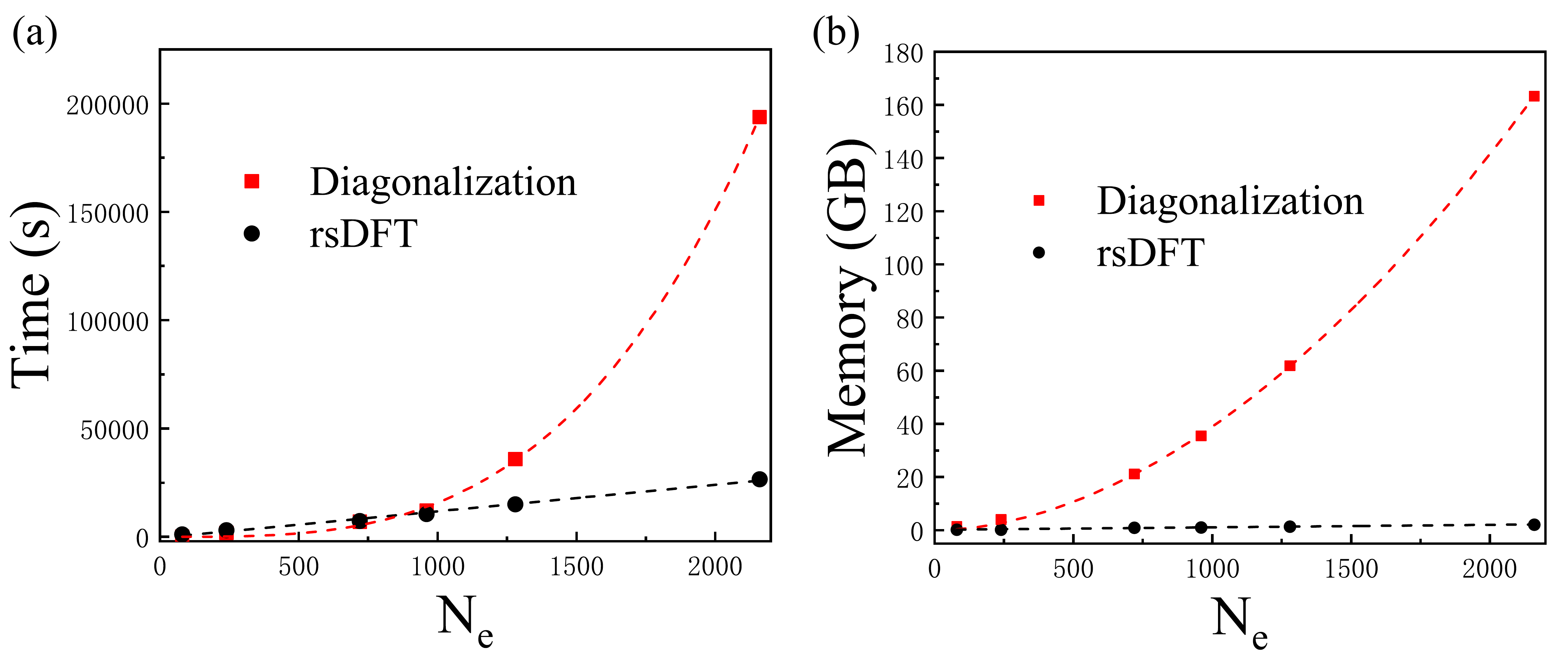} \caption{ Time cost of per iteration (a) and memory cost (b) during self-consistent calculations for fullerenes with different numbers of electrons. Black points refer to the traditional KS-DFT method with diagonalization, and red points refer to the rsDFT method without diagonalization. 
In rsDFT, we used 36 random samples for the average in the DOS and charge density calculations, the time step is $\tau=64\pi$ and the number of time steps is $N_{t}=36$.}
\label{fig:time} 
\end{figure}

\bibliography{references}